# THE INFLUENCE OF SUPERSHELLS AND GALACTIC OUTFLOWS ON THE ESCAPE OF IONIZING RADIATION FROM DWARF STARBURST GALAXIES

Akimi Fujita,[1,2,3] Crystal L. Martin,[3,4,5] Mordecai-Mark Mac Low,[2] and Tom Abel[6]

Received 2002 August 14; accepted 2003 August 26

## ABSTRACT

We study the escape of Lyman continuum radiation from the disks of dwarf starburst galaxies, with and without supershells, by solving the radiation transfer problem of stellar radiation through them. We model disks with $M_d = 10^8$–$10^{10}\, M_\odot$, with exponential surface density profiles as a function of redshift, and model the effects of repeated supernova explosions driving supershells out of the disks, using the hydrodynamic simulation code ZEUS-3D. The amount of star formation is assumed proportional to mass above some density threshold. We vary the threshold to explore the range of star formation efficiencies, $f_* = 0.006$, $0.06$, and $0.6$. We find that the interstellar gas swept up in dense supershells can effectively trap the ionizing photons before the supershells blow out of the disks. The blowouts then create galactic outflows, chimneys that allow the photons to escape directly to the intergalactic medium. Our results are consistent with escape fractions of less than 0.1 measured in local dwarf starburst galaxies, because they are likely observed while the starbursts are young, before blowout. We suggest that high-redshift dwarf starburst galaxies may make a substantial contribution to the UV background radiation with total escape fractions $\gtrsim 0.2$, as expected if star formation efficiencies are $\gtrsim 0.06$.

*Subject headings:* galaxies: dwarf — galaxies: formation — galaxies: starburst — H II regions — ISM: bubbles

## 1. INTRODUCTION

The standard big bang theory predicts that the primordial gas recombines around $z \sim 1100$. However, the absence of a Gunn-Peterson trough in quasar spectra (Gunn & Peterson 1965; Becker et al. 2001; Djorgovski et al. 2001) indicates that the intergalactic medium (IGM) is highly ionized by redshift $z \sim 5$ by an ionizing background composed of Lyman continuum radiation from quasars (accretion-powered massive black holes) and galaxies (young massive stars). The relative mixture of these two components as a function of redshift is poorly constrained by observations. It is important to understand the nature of the background radiation and its cosmic history because most baryons reside in the IGM, where stars and galaxies are born.

The contribution of ultraviolet (UV) background radiation from quasars has been studied in a number of papers (e.g., Madau, Haardt, & Rees 1999; Donahue & Shull 1987; Shapiro & Giroux 1987; Bajtlik, Duncan, & Ostriker 1988), but quasars alone appear inadequate to produce enough UV radiation to reionize the IGM at high redshift because their number density declines steeply above $z = 3$ (e.g., Fan et al. 2001). On the other hand, a large number of star-forming galaxies at $2 \lesssim z \lesssim 4$ have recently been discovered using the Lyman break technique (Steidel et al. 1996), and the inferred star formation rate is found to stay constant from $z = 1$ up to at least $\sim 4$ (Steidel et al. 1999). This implies that the contribution from galaxies increases in importance with redshift and must have dominated when the first stars and galaxies were forming (see, however, Haiman & Loeb 1998 for the possible roles of obscured or low-luminosity high-redshift quasars). It is crucial to understand the fraction of UV photons that can escape from galaxies.

Determining the escape fraction of ionizing photons from galaxies is no easy task, so there are only a limited number of observational estimates available. The escape fractions of Lyman continuum photons in four nearby starburst galaxies were first measured by Leitherer et al. (1995) using the Hopkins Ultraviolet Telescope. They observed flux around 900 Å, which provides a reliable estimate of the total number of Lyman continuum photons, even if starburst parameters are uncertain. The sample was reevaluated by Hurwitz, Jelinsky, & Dixon (1997), who found upper limits on the escape fraction of $\sim 0.05$–$0.1$. The only exception is Mrk 66, which had an upper limit of $\sim 0.57$. The sample was biased toward having the most favorable conditions for a high escape fraction of UV photons (e.g., high H$\alpha$ and UV luminosities but low infrared luminosities). Heckman et al. (2001) used the *Far-Ultraviolet Spectroscopic Explorer* (*FUSE*; Moos et al. 2000) to study the far-UV spectra of six nearby galaxies, three of which are dwarf starburst galaxies. They found that the strong C II $\lambda 1036$ interstellar absorption line is essentially black in all the galaxies. Since the opacity of the neutral interstellar medium (ISM) below the Lyman edge will be significantly larger than in the C II line, they estimated the escape fractions of Lyman continuum photons to be less than 0.06. Deharveng et al. (2001) recently constrained the escape fraction of Lyman continuum radiation from a starburst galaxy, Mrk 54, to be less than 0.062, using *FUSE* and following the same method as Leitherer et al. (1995) and Hurwitz et al. (1997).

Steidel, Pettini, & Adelberger (2001) studied Lyman continuum emission from the composite spectrum of 29 Lyman break galaxies (LBGs) at $z \sim 3.4$. The ratio of emergent flux density at rest-frame 1500 Å to that in the Lyman continuum, $L(1500)/L(900) = 4.6 \pm 1.0$, implies that the escape fraction is very high, greater than 0.5. The ratios $L(1500)/L(900)$ calculated for the nearby starburst galaxies still suggest that the escape is less easy from nearby galaxies than from the LBGs (Deharveng et al. 2001). The estimated comoving emissivity from LBGs, $(1.2 \pm 0.3) \times 10^{26}~h$ ergs s$^{-1}$ Hz$^{-1}$ Mpc$^{-3}$ (Einstein–de Sitter universe with the Hubble constant $H_0 = 100~h$ km s$^{-1}$ Mpc$^{-1}$), exceeds that from QSOs by a factor of about 5 at $z \sim 3$. However, their sample is drawn from the bluest quartile of intrinsic far-UV colors, which suggests that it may be biased toward large escape fractions and UV luminosities. Very recently, Fernandez-Soto, Lanzetta, & Chen (2003) studied the UV escape fractions in high-redshift galaxies by using *Hubble Space Telescope* images and found that no more than 0.1 of Lyman continuum photons escape from high-redshift galaxies.

[1] Department of Astronomy, Columbia University, 550 West 120th Street, New York, NY 10027; fujita@physics.ucsb.edu.
[2] Department of Astrophysics, American Museum of Natural History, Central Park West at 79th Street, New York, NY 10024; mordecai@amnh.org.
[3] Department of Physics, University of California, Santa Barbara, CA 93109; cmartin@physics.ucsb.edu.
[4] Packard Fellow.
[5] Alfred P. Sloan Foundation Fellow.
[6] Department of Astronomy, Pennsylvania State University, State College, PA 16802; hi@tomabel.com.

The observations seem to suggest that only a small fraction of ionizing radiation can escape from galaxies, and there is a possibility that more ionizing radiation escapes from galaxies at higher redshift. This result appears to contradict the theoretical predictions on the escape fractions from starburst galaxies (Ricotti & Shull 2000; Wood & Loeb 2000). Ricotti & Shull (2000) studied the escape fractions of ionizing radiation from small spherical high-redshift galaxies with dark matter masses of $M_{\rm DM} = 10^6$–$10^8~M_\odot$ and found the escape fractions to be less than 0.1 in halos with $M_{\rm DM} \gtrsim 10^7~M_\odot$. Wood & Loeb (2000) studied both low- and high-redshift galaxies and miniquasars with well-formed disks having $M_{\rm DM} = 10^9$–$10^{12}~M_\odot$. Their model galaxies have ionizing sources distributed throughout the disks as $\propto n$ or $n^2$, and their model miniquasars have single ionizing sources at the center of the disks; the miniquasars have higher total ionizing luminosities than the galaxies. The escape fractions of Lyman continuum photons are found to be $\sim 1$ at low redshift for all the galaxies and miniquasars and decrease to less than 0.01 for the galaxies but only to $\sim 0.3$ for the miniquasars at $z \sim 10$. Therefore, they suggest that the reionization must have been dominated by miniquasars. The study by Benson et al. (2001) supports the above results.

The escape fraction of ionizing photons from our Galaxy studied by Dove & Shull (1994) and Dove, Shull, & Ferrara (2000) is consistent with that required to sustain the Reynolds layer. Dove & Shull (1994) studied the escape fraction from a collection of OB associations distributed throughout the Galactic disk and found it to be $\sim 0.1$. The studies by Ciardi, Bianchi, & Ferrara (2002) and Clarke & Oey (2002) confirm the above result. Dove et al. (2000) extended the study to include the dynamics of bubbles and H II chimneys created by OB associations and found that the shells of expanding superbubbles quickly trap the ionizing flux. The estimated escape fraction decreases roughly by a factor of 2.

Although the high-redshift observations remain very preliminary, the low-redshift results seem to be converging toward a low escape fraction of ionizing radiation, less than 10%, not only from galaxies such as the Milky Way but from starburst galaxies. The discrepancy between the observations and the theoretical predictions of starburst galaxies may be because the gasdynamics driven by starbursts are not considered in the theoretical models. In this paper, we study the impact of supershells and galactic winds on the escape of ionizing radiation from dwarf starburst galaxies. We focus on dwarf galaxies because they are relatively simple systems. In addition, small-scale structures such as dwarf galaxies are the first ones to collapse in the early universe, according to the theory of hierarchical galaxy formation, and therefore may make a significant contribution to the reionization of the universe.

To study the escape fraction of ionizing radiation from dwarf starburst galaxies, we model disks with baryonic masses $M_d = 10^8$, $10^9$, and $10^{10}~M_\odot$ at redshifts $z = 0$, 3, and 8. Our motivation is to study the impact of supershells and galactic winds due to starbursts on the escape of ionizing radiation by using the Mac Low & Ferrara model (1999; hereafter MF99), combined with the photoionization code of Abel, Norman, & Madau (1999). We assume disks that are rotationally supported with exponential surface density profiles, on the basis of the predictions of standard hierarchical cosmologies (Mo, Mao, & White 1998). We note that the presence of a well-formed disk is uncertain at high redshift (e.g., Abel et al. 1998; Abel, Bryan, & Norman 2000) but argue that such idealized disks yield a lower limit to the escape fraction of ionizing radiation from a halo with given baryonic mass. In addition, our model is limited to a single ionizing source (i.e., starburst clump or star cluster) at the center of the disk, but we argue that centrally located sources also yield a lower limit to the escape fraction. In § 2, we describe our photoionization model with Lyman continuum photon luminosities, based on the Starburst99 model (Leitherer et al. 1999). In § 3, we describe our disk and halo models and how we model star formation in the disks. In § 4, we summarize our numerical methods, developed to model the impact of repeated supernovae (SNe) in the galactic disks, based on MF99. We show results on the escape fractions from undisturbed disks without supershells in § 5.1 and those from disks disturbed by starburst bubbles with supershells and outflows in § 5.2. In § 6, we compare our model results with observations of dwarf starburst galaxies and also attempt to compute the contribution of dwarf starburst galaxies to the UV background radiation at high redshift. Conclusions follow in § 7.

## 2. PHOTOIONIZATION MODEL

We compute the propagation of ionization fronts by using a photon-conserving radiative transfer code, developed to simulate inhomogenous reionization self-consistently within a cosmological hydrodynamical simulation (Abel et al. 1999). It was slightly modified for our problem. The code solves for the time at which the ionization front (I-front) reaches each grid cell along many different rays by solving the Strömgren equation along each ray:

$$4\pi r_{\rm I}^2 \frac{dr_{\rm I}}{dt} n_{\rm H\,I} = N_{\rm ph}(t) - 4\pi\alpha_{\rm B}\int_0^{r_{\rm I}} n_p n_e r^2\, dr\,, \qquad (1)$$

where $r_{\rm I}$ is the radial distance of the I-front at that angle from an ionizing source, $n_{\rm H\,I}$ is the number density of neutral hydrogen, $n_p$ and $n_e$ are number densities of protons and electrons, $N_{\rm ph}(t)$ is a photon luminosity, and $\alpha_{\rm B}$ is the case B recombination constant. The on-the-spot approximation is used to treat the effects of the diffuse emission of ionizing photons. This is expected to be reliable for scenarios in which stellar sources are responsible for the ionization of hydrogen (Abel et al. 1999).

The code solves for the radiative transfer around a point source; in our study this is a central starburst source. We choose a single site of star formation for numerical simplicity but also based on the assumption that some sort of dynamical instability, accompanied by the transfer of angular momentum, drives gas to the center. However, local observations of dwarf galaxies show that the starbursts occur in isolated clumps in the disks where the column densities of gas are greater than $10^{21}$ cm$^{-2}$ (Skillman 1997), so our assumption of a central source can certainly be improved. Another way of looking at the problem is that we do not actually calculate the escape fraction of ionizing radiation from a realistic starburst galaxy but rather from a single star-forming clump or star cluster in the galaxy. We later argue in § 3 that we are computing the lower limit for the escape fraction for the single clump by placing it at the center and that the result with a point source can be extended to a case in which there are multiple ionizing sources scattered around the disk.

We consider two different starburst scenarios: (1) an instantaneous starburst and (2) multiple starbursts spread over a time of 20 Myr. The photon luminosity below the Lyman limit of the cluster is taken from the Starburst 99 model (Leitherer et al. 1999), as a function of time. The model is based on a power-law initial mass function (IMF) with exponent $\alpha = 2.35$ between low-mass and high-mass cutoff masses of $M_{\rm low} = 1\, M_\odot$ and $M_{\rm up} = 100\, M_\odot$, respectively, with metallicities ranging from $\mathscr{Z} = 0.001$ to $0.04$. This approximates the classical Salpeter (1955) IMF. Most observations of star-forming and starburst regions are consistent with the Salpeter IMF, although uncertainties can be large (Leitherer et al. 1999). In Figure 1, we plot the photon luminosity as a function of time $N_{\rm ph}(t)$ for three different models in which $10^6\, M_\odot$ of stars between 1 and 100 $M_\odot$ form, producing $\sim 40{,}000$ SNe. Note that with the instantaneous-burst model, most photons are produced in the first few megayears of the burst, during the lifetime of the most massive stars. For different strengths of starburst, we linearly scale the above luminosity, $N_{\rm ph}(t, M = 10^6\, M_\odot)$, according to the mass available for the starburst, $M_{\rm SB}$, as follows: $N_{\rm ph}(t, M_{\rm SB}) = N_{\rm ph}(t, 10^6\, M_\odot) M_{\rm SB}/(\xi \times 10^6\, M_\odot)$, with the coefficient, $\xi = 2$, which accounts for the mass of stars below 1 $M_\odot$ with the power-law turnover below $M = 0.1\, M_\odot$. The number of photons produced per solar mass is $5.8 \times 10^{60}$.

The radiation transfer code is a postprocessing step that operates on a given density distribution at a given time step in our simulation. Therefore, the time evolution of bubble structures (see § 4) does not include ionization. However, the additional pressure due to photoionization is small compared with the bubble internal pressure. We can accurately compute the instantaneous escape fraction $f_{\rm esc}(t)$ (see § 5.1 for the definition) at a given time, since the ionization front propagates sufficiently rapidly to adjust almost instantaneously to a changing density distribution.

## 3. DISK

The escape fraction is very sensitive to the H I distribution in a galactic disk. In our study, we assume that there is a single ionizing source at the center of the galactic disk. In this case, the shortest axisymmetric path for the photons to escape is in the vertical direction. With the same amount of gas and a fixed column density distribution from the midplane, $N_{\rm H\,I}(R) = \int_0^\infty n_{\rm H\,I}(R, Z) dZ$ in axisymmetric cylindrical coordinates, it is harder for ionizing photons to escape in the vertical direction from a thick disk with a large scale height $H$ than from a thin disk with a small $H$. This is because the gas located high above the midplane at distance $Z$ can trap the photons reaching that height, which are diluted by the inverse square law, $1/4\pi Z^2$, more efficiently than the gas located close to the midplane, which encounters a larger photon flux, despite the higher density closer to the midplane. We show later in this section that the formation of shells of swept-up ISM by superbubbles complicates this situation.

To show that it is harder for photons to escape a thick disk, it is useful to define the effective recombination rate in a given direction through a disk:

$$\Xi = \int_0^\infty \alpha_{\rm B} n_p n_e r^2\, dr \;, \qquad (2)$$

where $r = (R^2 + Z^2)^{1/2}$ is the distance from the central ionizing source located at $r = 0$. The photon luminosity per steradian along each ray at a given time, $N_{\rm ph}(t)/4\pi$, must exceed $\Xi$ for any photon to escape in the given direction.

For example, if we assume the vertical distribution of neutral hydrogen gas is Gaussian with number density $n(Z) = n_* \exp(-Z^2/H^2)$ or exponential with $n(Z) = n_* \exp(-Z/H)$, the radial column density from the ionizing source at $R = 0$ is $N_{\rm H\,I,0} = N_{\rm H\,I}(R=0) \approx (\sqrt{\pi}/2) n_* H$ (Gaussian) or $n_* H$ (exponential), so the effective recombination rate in the $Z$ direction is

$$\Xi_Z = \begin{cases} \alpha_{\rm B} n_*^2 H^3 \left(\dfrac{1}{8}\sqrt{\dfrac{\pi}{2}}\right) & \text{for Gaussian} \\ \alpha_{\rm B} n_*^2 H^3 \left(\dfrac{1}{4}\right) & \text{for exponential} \end{cases} \propto N_{\rm H\,I,0}^2 H$$

$$(3)$$

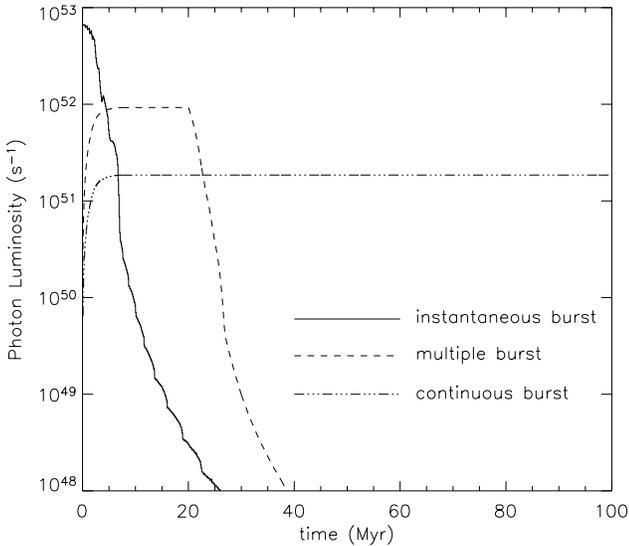

Fig. 1.—Number of Lyman continuum photons produced per second as a function of time $N_{\rm ph}(t)$. Three different models are shown: the instantaneous-burst model (solid line), multiple-burst model convolved with a timescale of 20 Myr (dashed line), and constant star formation model with a duration of 100 Myr (triple-dot–dashed line).

When $N_{\rm H{\scriptsize I},0}$ is fixed, a larger scale height $H$ means a larger effective recombination rate $\Xi_Z$.

On the other hand, if we assume that the radial distribution of gas is exponential with number density $n(R) = n_* \exp(-R/R_d)$, where $R_d$ is the scale radius for column density distribution, then the effective recombination rate in the $R$ direction is

$$\Xi_R \propto N_{\rm H{\scriptsize I},0}^2 H^{-2} R_d^3 . \quad (4)$$

When $N_{\rm H{\scriptsize I},0}$ is fixed, a large $H$ means a smaller effective recombination rate, $\Xi_R$. It is easier for the photons to move in the radial direction in a thick disk with a larger scale height, $H$, than in a thin disk with a smaller $H$. Note that $\Xi_R > \Xi_Z$ as long as $R_d > H$, although the radial density profile in our model disk falls off faster than exponential. In the next subsection, we constrain $R_d$ and $N_{\rm H{\scriptsize I}}(R)$ by a standard theory of hierarchical galaxy formation.

### 3.1. Disk and Halo Model

We set up a disk with an exponential surface density profile, $\Sigma_g = \Sigma_0 \exp(-R/R_d)$, where $R_d$ is a scale radius, consistent with observations of local compact dwarf galaxies (van Zee, Skillman, & Salzer 1998). To model a disk with an exponential surface density profile, we follow Mo et al. (1998), who compute the scale radius $R_d$ by assuming that a fraction of halo mass, $m_d = M_d/M_h$, settles into a non–self-gravitating, rotationally supported disk in a singular isothermal halo with a fraction of halo angular momentum, $j_d = J_d/J_h$. The halo angular momentum is produced by tidal torques from the background density distribution. The conventional way to specify the angular momentum is by the dimensionless spin parameter, $\lambda = J_h |E| M_h^{-5/2}$, where $J_h$ is the total angular momentum of the halo and $E$ is its binding energy. $N$-body experiments show that the distribution of $\lambda$ is Gaussian, with a peak at $\lambda \simeq 0.05$ (e.g., Cole & Lacey 1996).

If gas with given angular momentum settles into an exponential disk, the scale radius is given by

$$R_d = \frac{1}{\sqrt{2}} \left(\frac{j_d}{m_d}\right) \lambda r_{200} , \quad (5)$$

where $r_{200}$ is the virial radius of a halo, defined as the radius at which the mean overdensity in the halo is $200\rho_{\rm bg}$ and $\rho_{\rm bg}$ is the background density. The central surface density is then given by $\Sigma_0 = M_d/2\pi R_d^2$, which enables us to fully specify the shape of the disk. This method provides a good fit to the observed size distribution of galactic disks with $j_d/m_d = 1$ (Mo 1998).

Equation (5) is derived with an isothermal profile for the halo,

$$\rho_h(r) = \frac{V_c^2}{4\pi G r^2} , \quad (6)$$

despite the unphysical singularity at its center, since with this profile the total binding energy $E$ can easily be calculated analytically. The central structure does not markedly influence $R_d$. Once we have defined $R_d$ using this approximation, though, we use a softened isothermal halo with a central core to specify other halo parameters and to solve for the galactic structure in hydrostatic equilibrium. We use the form given by Salucci & Persic (1997) based on observations of dwarf galaxies,

$$\rho_h(r) = \frac{\rho_0}{1 + (r/r_0)^2} , \quad (7)$$

where $r_0$ and $\rho_0$ are the core radius and the central density (Binney & Tremaine 1987). The virial radius is then

$$r_h \equiv r_{200} = \left(\frac{3\rho_0}{200\rho_{\rm bg}}\right)^{1/2} r_0 . \quad (8)$$

As in MF99, we calculate $r_0$ and $\rho_0$ as a function of halo mass based on the empirical relation between $r_0$ and $\rho_0$ for nearby dwarf galaxies (Burkert 1995) extrapolated to the early universe by using the usual cosmological scaling relations,

$$r_0 = (8.9 \times 10^{-6} \text{ kpc}) \left(\frac{M_h}{1\, M_\odot}\right)^{1/2}$$
$$\times h^{1/2} \left[\frac{\Omega_0}{\Omega(t)}\right]^{-1/3} (1+z)^{-1} , \quad (9)$$

$$\rho_0 = (6.3 \times 10^{10}\, M_\odot \text{ kpc}^{-3}) \left(\frac{M_h}{1\, M_\odot}\right)^{-1/3}$$
$$\times h^{-1/3} \left[\frac{\Omega_0}{\Omega(t)}\right] (1+z)^3 . \quad (10)$$

We extended the relations to high-redshift galaxies by assuming that the ratio of the core radius to the virial radius and the ratio of the core density to the background density stay constant.

The disks in our models are set up in hydrostatic equilibrium in the potentials of the dark matter halos and, in some models, the disks themselves. The gas that is important for trapping ionizing radiation is all assumed to be in the warm neutral state, heated and stirred up by star formation prior to the major starburst that we model, yielding an effective sound speed, including a turbulent contribution, of $c_s \sim 10$ km s$^{-1}$ with corresponding effective temperature $T \approx 10^{3.7}$ K. The situation is more complicated in a real multiphase gas disk. However, only a few SNe from massive stars suffice to blow through a thin disk of cold neutral medium similar to those observed (e.g., Mac Low & McCray 1988), thereby creating a funnel for ionizing radiation to escape. In addition, ionizing radiation escapes through a thin disk with a small scale height more easily than a thick disk with a large scale height. Therefore, we do not consider the thin cold disk in our calculations.

MF99 used the dark-to-baryonic mass ratio $\phi = M_h/M_g$ observed in local spiral galaxies, $\phi \simeq 34.7 M_{g,7}^{-0.29}$, where $M_{g,7} = M_g/10^7\, M_\odot$, hereafter $\phi_{\rm lcl}$, where "lcl" indicates "local" (Persic, Salucci, & Stel 1996). The relation was derived from a sample of larger galaxies with $M_h \gtrsim 10^{10}\, M_\odot$ in the local universe, but MF99 extrapolated the result to lower masses because the trend of increasing dark matter fraction with decreasing galaxy mass is indeed consistent with some observations (e.g., Mateo 1997). However, the extrapolation of this result to higher redshift is imperfect.

In Figure 2, we plot $M_h$ as a function of $M_g$ expected with the observed $\phi_{\rm lcl}$ relation and with the standard picture of the hierarchical $\Lambda$CDM cosmology by using $\Omega_0 = 0.37$ and $\Omega_b = 0.05$, which corresponds to $\phi = 7.4$. The difference appears in both lower and higher mass disks. Using the $\phi_{\rm lcl}$ relation underpredicts $M_g$ in lower mass halos and overpredicts it in higher mass halos, compared with the value of $M_g$ predicted in the $\Lambda$CDM model, in which baryons are assumed to simply follow the dark matter. The underprediction can be interpreted to mean that a significant amount of gas was depleted or blown away in the shallow potentials of small dwarf galaxies because of the past starburst activities, as seen in the simulations of MF99. The overprediction at high masses can be interpreted to mean that in larger galaxies the deeper potentials retain more gas in the halos and the cooling is also more efficient ($\propto n^2$). However, we show in § 5.1.5 that the difference in $M_g$ between $\phi_{lcl}$ and $\phi = 7.4$ is too small to influence the escape fractions of ionizing photons in any of the halos we examine.

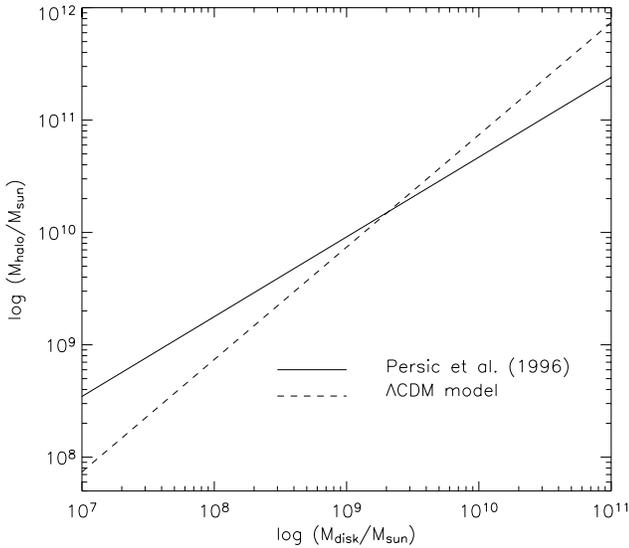

FIG. 2.—Mass of dark matter halos as a function of the mass of baryonic matter in the halos, with the $\phi_{lcl}$ relation (Persic, Salucci, & Stel 1996) and $\phi = 7.4$ expected with $\Lambda$CDM model ($\Omega_0 = 0.37$ and $\Omega_b = 0.05$), where $\phi = M_h/M_d$.

The scale heights of the disks are determined by setting the gas into hydrostatic equilibrium in the halo and, in some cases, the disk potentials. When a disk potential is included, we use the thin-disk approximation (Toomre 1963). The additional disk potential makes a disk thinner than otherwise. We refer to the disks with both dark matter and disk potentials as "thin" disks and those with only dark matter potentials as "thick" disks. We list the scale heights of our model disks in Table 1. The vertical gas distributions are roughly exponential in the thin disks but Gaussian in the thick disks. The presence of disk potentials makes the disks unreasonably thin when large surface densities are expected at high redshift and in large galaxies. For example, a thin disk with $M_d = 10^{10}~M_\odot$ at $z = 8$ has a scale height $H$ of less than 10 pc. Therefore, we consider our thick disks to be more appropriate to describe high-redshift disks, especially when highly dynamical processes of cooling and heating are expected after halo formation. A thin disk with $M_d = 10^{10}~M_\odot$ at $z = 0$ has only $H = 19$ pc, but a thick disk has $H = 980$ pc, which is unreasonably large. Therefore, we experiment with both thick and thin disks as the opposite extrema of possible gas distributions within the observational and theoretical limits.

We do not consider preexisting clumpiness of the disk, but we do consider the extreme case of clumpiness: shell formation due to starburst bubbles. If shells of swept-up ISM are formed at $r_{sh}$ with average density $n_{sh}$ and thickness $\Delta r_{sh}$, the column density from the central ionizing source is $N_{sh} = n_{sh}\Delta r_{sh}$ in spherical coordinates $[r = (Z^2 + R^2)^{1/2}]$ and the effective recombination rate is

$$\Xi_{sh} = \alpha_B n_{sh} N_{sh} r_{sh}^2 , \quad (11)$$

according to equation (2). Since the preferred direction for the growth of bubbles and therefore for the escape of photons is the vertical direction, we compare the effective recombination rate due to the shell, $\Xi_{sh} \propto n_{sh} N_{sh} Z_{sh}^2$, with that in an undisturbed disk, $\Xi_Z \propto N_{H1,0}^2 H$, from equation (3) in the vertical direction. We assume a Gaussian or exponential distribution of hydrogen as before.

We can assume that the shell, at least initially, is bounded by an isothermal shock because the swept-up gas in the shells cools very effectively in our model galaxies, as we show in § 5.2. In the absence of magnetic fields, the isothermal shock compression is $n_{sh} = \mathcal{M}^2 n(Z_{sh})$. A perpendicular magnetic field, however, reduces the compression behind an isothermal shock to $n_{sh} = \sqrt{2}\mathcal{M}_A n(Z_{sh})$, where $\mathcal{M}_A = v(4\pi\rho)^{1/2}B^{-1}$ is the Alfvénic Mach number of the shock. In the case of a Gaussian atmosphere, $n(Z_{sh}) = n_* \exp(-Z_{sh}^2/H^2)$, while in an exponential atmosphere $n(Z_{sh}) = n_* \exp(-Z_{sh}/H)$, where $n_*$ is the central number density. The ratio of the effective recombination rate due to the shell to that in an undisturbed disk is

$$\Xi_{sh}/\Xi_Z = \begin{cases} 2\sqrt{2} n_{sh}(Z_{sh}) & \text{for Gaussian} \\ 8 n_{sh}(Z_{sh}) & \text{for exponential} \end{cases} \quad (12)$$

in the limit of $Z_{sh} > H$. The shells trap the ionizing photons more effectively than the undisturbed disk, as long as $Z_{sh}/H$ is less than $[\log(2\sqrt{2}\mathcal{M}^2)]^{1/2}$ or $\log(8\mathcal{M}^2)$, respectively, or in the magnetized case $(\log 4\mathcal{M}_A)^{1/2}$ or $\log(8\sqrt{2}\mathcal{M}_A)$, respectively. For example, $\Xi_{sh}$ is greater than $\Xi_Z$ if the bubble is within $2.4H$ in a thick disk ($H = 580$ pc) and within $7.0H$ in a thin disk ($H = 100$ pc) with $M = 10^8~M_\odot$ at $z = 0$, with $\mathcal{M} = 10$.

In addition, the starburst-driven shells can trap the radiation more effectively than numerous clumps distributed throughout the disk, because such clumps would leave only a diluted ISM and allow for a larger escape fraction (Wood & Loeb 2000).

Our standard model applies a cold dark matter (CDM) cosmology with $\Omega_0 = 1$ and $h = 0.5$ and the observed dark-to-baryon ratio $\phi_{lcl}$, as in MF99, and spin parameter $\lambda = 0.05$. We experiment with both thin and thick disks, that is, both including and neglecting disk potentials. We also show in § 5.1.5 that changing the cosmology to the standard $\Lambda$CDM ($\Omega_0 = 0.37$, $\Omega_\Lambda = 0.63$, $h = 0.7$, $\Omega_b = 0.05$) does not influence the escape fractions.

TABLE 1
SCALE HEIGHTS OF THIN AND THICK DISKS

| $M_d$ ($M_\odot$) | Thin Disk, $z = 0$ (pc) | Thick Disk, $z = 0$ (pc) | $z = 3$ (pc) | $z = 8$ (pc) |
|---|---|---|---|---|
| $10^8$ | 100 | 580 | 71 | 21 |
| $10^9$ | 56 | 750 | 93 | 27 |
| $10^{10}$ | 19 | 980 | 121 | 36 |

There are limitations to our study. We model idealized rotationally supported disks with uniform single-phase ISM and exponential surface density profiles. Our model disks approximate the dwarf galactic disks observed in the local universe reasonably well, but we do not know whether such disks initially form at high redshift after halos collapse. A realistic scenario is that the stars form as the gas cools and collapses then explode as SNe and stir up the rest of the cooled gas in the halos. Disks may not form at all before star formation can occur, according to the study of the formation of first stars with high-resolution three-dimensional adaptive mesh refinement simulations by Abel et al. (1998, 2000). We think that a more turbulent dynamical ISM than the one we study will allow more photons to escape through funnels of hot gas. In this study, we solve for the lower limits of escape fractions by examining well-formed rotationally supported disks at all redshifts.

### 3.2. Modeling the Star Formation Rate

We considered several scenarios for the star formation in our disks since few empirical constraints are available for high-redshift galaxies. These scenarios share some common characteristics. First, we use a threshold density to delimit the star-forming region of the disk. This constraint effectively differentiates the star formation in systems with different angular momentum. Second, we assume that the amount of star formation is proportional to the amount of gas in this region. And, finally, merger-driven star formation is recognized by considering gas distributions that are more centrally concentrated than our fiducial disk model. We define a star formation (or starburst) efficiency parameter $f_*$ and define the amount of gas available for a starburst as $M_{SB} = f_* M_d$.

The density threshold for star formation in nearby spiral galaxies (Kennicutt 1989) and starburst nuclei (Kennicutt 1998) is well described by models for the onset of gravitational instability (Kennicutt 1989, 1998). The critical surface density for star formation is determined by a column density cutoff based on Toomre's criterion, $\Sigma_c = 0.5\kappa c_s/\pi G$, where $\kappa$ is the epicyclic frequency. The coefficient 0.5, hereafter the gravitational threshold, is fitted to the observations by Martin & Kennicutt (2001). We assumed velocity dispersions maintained by turbulence, $c_s = 10$ km s$^{-1}$, to get the highest efficiency in trapping radiation and so to yield a lower limit to the escape fractions. This assumption is consistent with the values of velocity dispersions measured in nearby galaxies (van der Kruit & Shostak 1984; Dickey, Hanson, & Helou 1990; van Zee et al. 1998) but is uncertain at high redshift. It is also not clear that the same relation holds in dwarf galaxies (Hunter, Elmegreen, & Baker 1998). To explore other cutoffs, we also considered a constant column density cutoff, $\Sigma_c = 10^{21}$ cm$^{-2}$ (Skillman 1997).

The critical masses of our disks with $M_d = 10^8\ M_\odot$ at $z = 0, 3$, and 8 are $f_* M_d$ with $f_* = 0.0056, 0.097$, and $0.32$, respectively, by the gravitational threshold and $f_* = 0.013, 0.79$, and $0.94$, respectively, by the $10^{21}$ cm$^{-2}$ cutoff. Our standard model for the gas distribution produces fairly low star formation rates in the disk with $M_d = 10^8\ M_\odot$ at $z = 0$. The observations of blue compact galaxies (BCDs) by van Zee et al. (1998) show that there is another population of dwarf galaxies with higher central column densities and these tend to be the bursting systems most likely to contribute to escaping UV photons. Therefore, we increased the central gas density in our model disk with $M_d = 10^8\ M_\odot$ at $z = 0$ by decreasing the spin parameter $\lambda$ to 0.035 (within 1 $\sigma$ deviation of the mean, $\bar{\lambda} = 0.05$). The new critical mass is $f_* M_d$ with $f_* = 0.074$ with the gravitational threshold and $f_* = 0.2$ with the $10^{21}$ cm$^{-2}$ cutoff. Figure 3 shows the surface density distributions of the disks with $M_d = 10^8\ M_\odot$ at $z = 0, 3$, and 8 with the critical surface densities predicted by the gravitational threshold and the $10^{21}$ cm$^{-2}$ cutoff. In this study, we consider three possible star formation efficiencies for single starburst clusters in our model disks: $f_* = 0.006, 0.06$, and 0.6. Note that $f_*$ also depends on the amount of metals present in dwarf galaxies, but the effects of metallicity will be smaller than the 2 orders-of-magnitude difference in $f_*$.

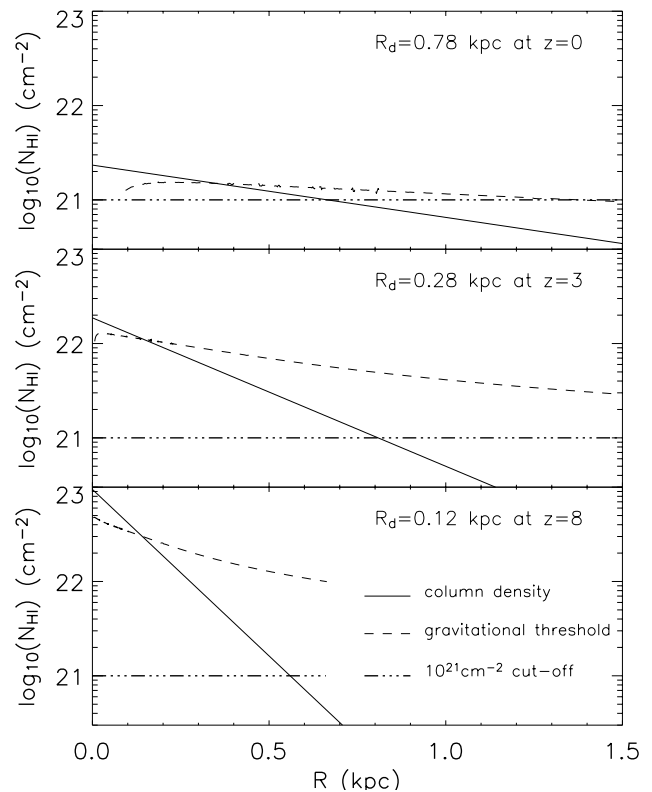

Fig. 3.—Surface density profiles for the disks with $M_d = 10^8\ M_\odot$ at $z = 0, 3$, and 8, compared with the critical surface densities predicted by the gravitational threshold and the $10^{21}$ cm$^{-2}$ cutoff. The sizes of the disks are $5.5R_d$ with $R_d \propto (1+z)^{-1}$. The critical surface density by the gravitational threshold is not drawn at the center because the gas is supported by pressure rather than rotation there.

Figure 4 shows the distributions of exponential scale radii $R_d$ and central surface densities $\Sigma_0$ for our model galaxies based on the standard model, compared with those for five BCDs by van Zee et al. (1998) and a number of late-type dwarf galaxies by Swaters et al. (2002). The scale radii and the surface densities of our model galaxies with $M_d = 10^8\ M_\odot$ and $\lambda = 0.035$ and with $M_d = 10^9\ M_\odot$ and $\lambda = 0.05$ at collapse redshift $z = 0$ are consistent with the observed distributions. We cannot in principle compare our higher redshift galaxies with the local BCDs and late-type dwarf galaxies. We also note that the CDM predictions of the disk spin parameter are very uncertain (Dalcanton, Spergel, & Summers 1997). A disk with a low spin parameter has a higher surface density and so a higher effective recombination rate: $\Xi_Z \propto N_{\rm H\,I,0}^2 \propto \lambda^{-4}$. However, the same disk yields a higher critical mass and so a higher star formation efficiency $f_* \propto \lambda^{-2.8}$ based on the Schmidt law (Schmidt 1959): $\dot{M}_* \propto \Sigma_0^{1.4}$. The increase in $\Xi_Z$ is compensated for by the number of ionizing photons $N_{\rm ph}$ due to the increase in $f_*$. Therefore, the qualitative results on the escape fractions of ionizing photons in a disk with a given baryonic mass can be applied to the same disk with a different spin parameter $\lambda$. Despite the uncertainty in $\lambda$ distribution, our models at $z = 0$ trace the upper boundary of the properties of local dwarf galaxies well (Fig. 4). The actual form of high-redshift disks remains the largest uncertainty, as we mentioned in the end of § 3.1.

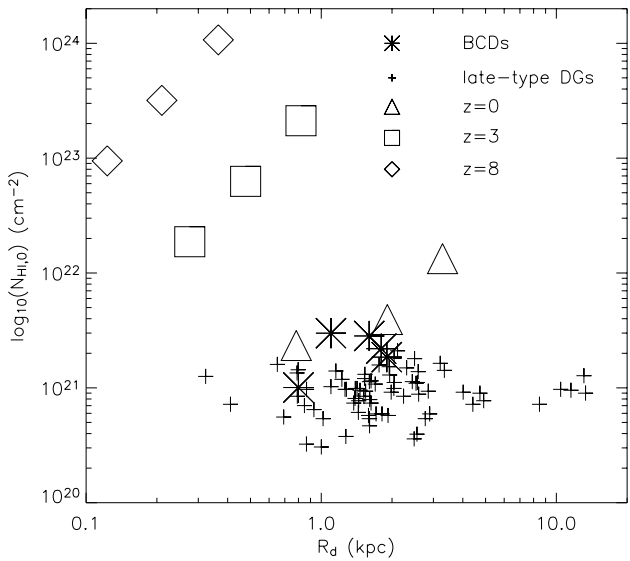

Fig. 4.—Distributions of exponential scale lengths $R_d$ and central surface densities $\Sigma_0$ in units of H I column density (cm$^{-2}$) of our model galaxies at $z = 0$ (*triangles*), $z = 3$ (*squares*), and $z = 8$ (*diamonds*) and those based on the observations of blue compact dwarf galaxies (BCDs; Fig. 11 from van Zee, Skillman, & Salzer 1998) and of late-type dwarf galaxies (Swaters et al. 2002).

A last limitation of our model is that we deal with only a single star-forming source at the center of a disk. In realistic starburst galaxies, we expect that multiple starburst clumps are scattered around the disks (e.g., Vacca 1996; Martin 1998). Recall that the photons find their way out of the disks first in the vertical direction ($\Xi_Z < \Xi_R$) and that $\Xi_Z \propto N_{\rm H\,I}^2(R)H(R)$ and $N_{\rm H\,I}(R) \propto \exp(-R/R_d)$, although $H$ increases as a function of $R$. This means that a clump placed at the center yields the lower limit for the escape fraction and the escape fraction we compute is a lower limit for the same starburst clump placed off center. A starburst clump forming above $Z = 0$ has a higher escape fraction on the side it forms. The lower limits of escape fractions from realistic galaxies can be inferred by summing the contributions from starburst clumps in the disks.

## 4. SHELL FORMATION AND GALACTIC OUTFLOWS

Large outflows of gas are observed from both dwarf and massive starburst galaxies in the local universe (e.g. Heckman et al. 2000; Martin 1999; Lehnert & Heckman 1996) and at high redshift (Franx et al. 1997; Pettini et al. 1998). These outflows are driven by the mechanical energy supplied by massive stars in the form of SNe and stellar winds (Tomisaka & Ikeuchi 1988; De Young & Heckman 1994; Suchkov et al. 1994, 1996; MF99; D'Ercole & Brighenti 1999; Silich & Tenorio-Tagle 1998, 2001). Deep H$\alpha$ imaging often shows filamentary or bubble-like structures in dwarf galaxies (Marlowe et al. 1995; Della Ceca et al. 1996; Martin 1998), which are thin dense shells of swept-up ISM. When there is sufficient time for the shells to cool and become dense, such cooled neutral shells of hydrogen may be able to trap the ionizing radiation inside the host halos.

Dove et al. (2000) studied the escape of ionizing radiation from the disk of the Milky Way and found that the shell of the superbubble driven by each OB association quickly traps or attenuates the ionizing flux from that association. The escape fraction of photons does not increase even after superbubbles of larger associations blow out of the H I disk and form dynamic chimneys because by that time the production of Lyman continuum photons has drastically fallen. However, the escape fraction was already very small, $\sim 0.1$, in their Galactic disk without these shells. This is because they modeled the disk as having OB associations with a power-law photon luminosity function, and the largest OB association has a photon luminosity of only $2 \times 10^{51}$ s$^{-1}$. Even in dwarf starburst galaxies, much higher photon luminosities are expected, greater than a few times $10^{52}$ s$^{-1}$, although they probably come from multiple star clusters (e.g., Martin 1998; see § 6.1). The collective power of stellar winds and tens of thousands of SNe in starburst galaxies can be much larger than that from any single OB association studied in their model. In our study, we examine how supershells and galactic outflows influence the escape of ionizing radiation from powerful starburst clumps.

To model the effects of shell formation due to repeated SN explosions from dwarf starburst galaxies, we use ZEUS-3D, a second-order Eulerian astrophysical gasdynamics code (Stone & Norman 1992; Clarke 1994) that uses Van Leer (1977) monotonic advection. Runs were done on Silicon Graphics Origin 2000 machines using eight processors and typically took $\sim$10–15 hr in the models at $z = 0$ for the computation up to 10 Myr and about 4 days in the models at $z = 8$ for computation up to 5 Myr after the onset of starbursts.

Our numerical methods and initial conditions generally follow MF99 (see their § 4): we use axisymmetric ratioed grids; a cooling curve by MacDonald & Bailey (1981), linearly scaled by a factor of 10 to account for low metallicity in dwarf galaxies; and a tracer field (Yabe & Xiao 1993) to avoid overcooling in the bubble interiors. Note that the cooling curve $\Lambda(T)$ in principle does not linearly scale; however, we also tested the cooling curve of Sutherland & Dopita (1993) with primordial gas composition. It agrees with the linearly scaled $\Lambda(T)$ by MacDonald & Bailey (1981) within a factor of a few in the temperature range between $10^5$ and $10^7$ K, which are the shock-heated temperatures of the shells expected in our simulations with $v_{sh} \approx 100$ km s$^{-1}$. We also find that the cooling of shells $\Lambda(T)n^2$ is not so sensitive to $\Lambda(T)$ but primarily to the density of the shells, $n$.

However, there are a few changes. As in MF99, the thermal energy source that drives a constant luminosity wind is set up as a source region with a radius of five zones. Since the sizes of galaxies decrease as $(1 + z)^{-1}$, the resolutions adopted also decrease roughly as $(1 + z)^{-1}$. This means that physical sizes of the source regions decrease at high redshift. We justify this by noting that smaller source regions favor shell formation. Here we test the maximum effect of shell formation on the escape of ionizing radiation, and smaller source regions give a firm lower limit.

The source region is set up initially in pressure equilibrium with the background, but at a temperature a factor $\eta$ larger than that of the background and with density correspondingly lower by a factor of $1/\eta$. The source region then has energy added to it at a rate $\dot{E} = L_{\mathrm{mech}}/V$ and mass at a rate $\dot{M} = \rho_0 \dot{E}/(\eta e_0)$, where $\rho_0$ and $e_0$ are the midplane mass and energy density of the background disk, respectively, to roughly keep the specific energy of the source region constant. The amount of mass in this wind accounts not only for SN ejecta but also mass that would have evaporated from the shells if we had included thermal conduction in our model. In fact, evaporation from the shells provides most of the mass within a realistic bubble (Weaver et al. 1977). MF99 fixed $\eta = 1000$, which was appropriate for the small range of mechanical luminosities $L_{\mathrm{mech}}$ that they studied. However, to take into account our broader range of $L_{\mathrm{mech}}$, we vary $\eta$ as a function of $L_{\mathrm{mech}}$ by directly equating the mass input rate $\dot{M}$ with the mass evaporation rate predicted by Weaver et al. (1997) for a spherical bubble,

$$\dot{M}_{\mathrm{ev}} \propto L_{\mathrm{mech}}^{27/35} t^{-4/35} \rho_0^{-2/35} . \quad (13)$$

As we have $t$ and $L_{\mathrm{mech}}$, we can estimate $\eta$ by noting the weak dependence of the mass evaporation rate on the density and choosing to set the density constant $n_0 \sim 10^3$ cm$^{-3}$. In this way, we keep the density and temperature within the bubble fairly close to the analytic solution of Weaver et al. (1977) and so can follow the interior dynamics with some degree of reliability.

We note that photoevaporation by stellar UV radiation from interior molecular clouds increases the amount of $10^4$ K gas within a superbubble. This will increase the surface area available for conductive flow, although it is hard to estimate how much will end up in the hot interior gas (McKee, van Buren, & Lazareff 1984; Shull & Saken 1995). However, within the range of mechanical luminosities considered in our study, $L_{\mathrm{mech}} \gtrsim 10^{39}$ ergs s$^{-1}$, the cooling due to the photoablated material of $\sim 10^4$ $M_\odot$ will not be important after the bubble grows greater than $\sim 30$ pc because of reduced interior density (Shull & Saken 1995).

We list the resolutions used for the computation of the escape fractions of ionizing radiation in undisturbed disks and in disks disturbed by galactic outflows in Table 2. We find that we need to resolve the scale heights of the undisturbed disks with more than 50 zones to get converged results with our photoionization code. For the outflow models, we are not able to fully resolve the shell formation when the shocks are isothermal with shell densities $n_{\mathrm{sh}} = n_d \mathcal{M}^2$ or $n_{\mathrm{sh}} = n_d \sqrt{2} \mathcal{M}_A$ in the presence of magnetic fields, where $n_d$ is the density of the ambient disk.

The cooling time of the shells formed by initially adiabatic shocks is $t_c = 3kT_s/n\Lambda(T)$, with postshock electron number density $n_e = n$. The shocked temperature is

$$T_s = (3\mu_s/16k)v_{\mathrm{sh}}^2 \simeq 1.3 \times 10^5 \left(\frac{v_{\mathrm{sh}}}{100 \text{ km s}^{-1}}\right)^2 \text{ K} , \quad (14)$$

where $\mu_s = (14/23)m_H$ is the mass per particle in the fully ionized postshock gas, assuming $n_{\mathrm{He}}/n_H = 0.1$, and $v_{\mathrm{sh}}$ is the shock velocity. To estimate the cooling time $t_c$, we use the analytic estimate of the Gaetz & Salpeter (1983) cooling function between $10^5$ and $10^7$ K from Mac Low & McCray (1988): $\Lambda(T) = (1.0 \times 10^{-23} \text{ ergs cm}^3 \text{ s}^{-1})\zeta T_6^{-0.7}$ with $T_6 = T/10^6$ K, where $\zeta$ is the metallicity relative to solar $\zeta = \mathcal{Z}/\mathcal{Z}_\odot$. Then we find that the instantaneous cooling time of the postshock gas in the shells is

$$t_c = 1.2 \times 10^4 \text{ yr} \left(\frac{v_{\mathrm{sh}}}{100 \text{ km s}^{-1}}\right)^{3.4} \left(\frac{n}{1 \text{ cm}^{-3}}\right)^{-1} \left(\frac{\zeta}{0.1}\right)^{-1} , \quad (15)$$

which remains short ($t_c \ll 1$ Myr), while the bubble is within a few scale heights of the disk, with the expected $v_{\mathrm{sh}} \lesssim 200$ km s$^{-1}$. Therefore, we expect isothermal shocks near the disks.

We show in Figure 5 an example of our resolution study of the behavior of isothermal shocks, for a thin disk with $M_d = 10^8$ $M_\odot$ at $z = 0$, with a mechanical luminosity $L_{\mathrm{mech}} = 10^{40}$ ergs s$^{-1}$. The expected density of swept-up shells behind unmagnetized isothermal shocks is $n_{\mathrm{sh}} = n_d \mathcal{M}^2$. The effective sound speed $c_s = 10$ km s$^{-1}$ in our simulations is driven by turbulence and photoionization, so the Mach number could be larger given the short timescales on which turbulence dissipates and hot stars disappear. Magnetic fields on the order of a microgauss may, on the other hand, reduce the compression in realistic galaxies to $n_{\mathrm{sh}} = n_d \sqrt{2} \mathcal{M}_A$, as much as an order of magnitude lower.

TABLE 2
Minimum Grid Resolutions Used in Modeling the Disks

| $M_d$ ($M_\odot$) | $z = 0$ (pc) | $z = 3$ (pc) | $z = 8$ (pc) |
|---|---|---|---|
| $10^8$ | 2.5 | 0.63 | 0.28 |
| $10^9$ | 4.4 | 1.1 | 0.49 |
| $10^{10}$ | 7.5 | 1.9 | 0.77 |

Note.—The minimum grid resolutions used in modeling the disks are roughly proportional to $R_{\mathrm{vir}} \propto M_h^{1/3}(1+z)^{-1}$. We selected all the disks at $z = 0$ and the disks with $M_d = 10^8$ $M_\odot$ at $z = 3$ and 8 for the bubble experiments (first row and first column). We also experimented with higher resolutions, 1.25 and 0.38 pc, for the model with $M_d = 10^8$ $M_\odot$ at $z = 0$ to test the sensitivity of shell cooling to the resolution.

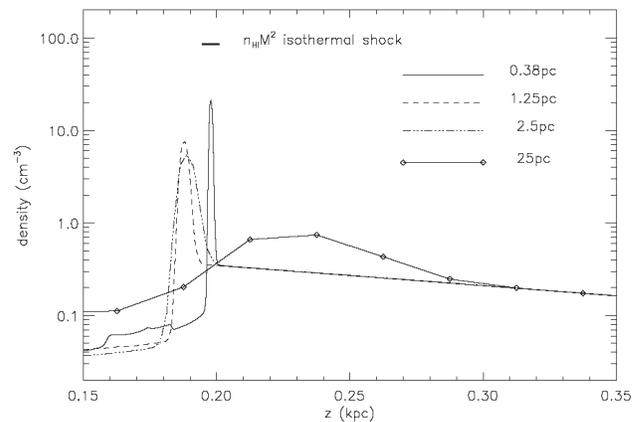

Fig. 5.—Resolution study for the disk with $M_d = 10^8$ $M_\odot$ at $z = 0$ with mechanical luminosity $L_{\mathrm{mech}} = 10^{40}$ ergs s$^{-1}$. The density of the shells is extremely sensitive to the resolution of the simulations. The theoretical value $n_d \mathcal{M}^2$ is indicated.

Figure 5 shows that the unmagnetized isothermal shell structure is better resolved as the resolution is increased from 25 to 0.38 pc, but the expected peak density of the shells due to isothermal shocks remains about a factor of 4 larger than the peaks in our simulations. At the same time, the fragmentation of the shells due to Rayleigh-Taylor instability is better resolved in the simulations with a higher resolution. The shells created by isothermal shocks would trap the ionizing radiation more effectively ($\Xi \propto n_{sh}$), but the fragmentation of the shells opens up holes through which the ionizing radiation can escape.

Note that the development of Rayleigh-Taylor instability in our model is limited not only by the resolution but also by the assumption of azimuthal symmetry. Mac Low, McCray, & Norman (1989) argued that the assumed symmetries of the problem largely determine the specific modes of the instability that are excited and thus the distribution of mass into particular fragments (e.g., rings in azimuthal symmetry) but that the appearance of unstable regions is independent of the symmetry assumed. We expect more and smaller shell fragments in three dimensions, but the escape of ionizing radiation is primarily determined by the opening angles of the chimneys created after bubbles blow out, as we show in § 5.2, so the suppressed modes of shell fragmentation within those angles will not substantially affect the outcome of our computations. We also do not expect thermal instability (Chevalier & Imamura 1982) or linear dynamical thin-shell instability (Vishniac 1983) to be important for shell fragmentation, based on the numerical simulations of these instabilities by Blondin & Cioffi (1989) and Mac Low & Norman (1993), respectively.

To estimate the actual ability of the shells to trap ionizing radiation in our limited resolution simulations, we set the shell density to the analytic value for an isothermal shock and compute the recombination measure $\Xi_{sh}$ at the positions of the shells that are not fragmented because of the instability. (This correction is needed only for disks at $z = 0$ because even unresolved shells in high-redshift disks are dense enough to trap the ionizing photons.) To make the correction, we set a temperature threshold $T_{th} = 5 \times 10^4$ K in our photoionization code, to identify cooled shells that have not yet fragmented. Then we make a correction $\Xi_{sh} = \alpha_B n_d \mathcal{M}^2 N_{sh} r_{sh}^2$ according to equation (11) if the maximum density in the identified shell is smaller than $n_d \mathcal{M}^2 \approx n_d [v_{sh}/(10\ km\ s^{-1})]^2$, the approximation applied within the uncertainties of magnetic fields and thermal properties of the shells. This approximation tends to overestimate the shell density at larger scale heights in low-redshift galaxies but is appropriate to test the maximum effect of the shells on the escape of ionizing radiation. We can reliably read $N_{sh}$ from the lower resolution simulations, since it is independent of cooling. Note that shell formation is more sensitive to the resolution than to the cooling function $\Lambda(T)$ applied with linear metallicity dependence because the cooling is proportional to $\Lambda(T) n^2$.

We assume mechanical luminosity to be constant in time for a given $f_*$ in our simulations, by taking the average of the luminosity $\mathscr{L}_{mech}(t)$ drawn from the Starburst 99 model (Leitherer et al. 1999). We apply $Z = 0.008$ at $z = 0$ based on the observations of local dwarf galaxies (Skillman, Kennicutt, & Hodge 1989; Hunter, Gallagher, & Rautenkranz 1982) but $Z = 0.001$ at high redshift, assuming that the timescale for metal replenishment by previous star formation prior to the major starbursts that we model is very short, much less than the Hubble time at a given redshift (Wada & Venkatesan 2003; A. Fujita et al. 2003). Figure 6 shows the evolution of $\mathscr{L}_{mech}(t)$ as a function of time when $10^6\ M_\odot$ of gas is converted to stars based on four different starburst scenarios. When the metallicity is fixed, the mechanical luminosity $\mathscr{L}_{mech}(t)$ is about an order of magnitude lower in the first few megayears with the multiple-burst model than with the instantaneous-burst model, although $\mathscr{L}_{mech}(t)$ soon rises to the same average mechanical luminosity expected with both star formation models. When the starburst model is fixed, $\mathscr{L}_{mech}(t)$ is about an order of magnitude lower in the first few megayears with $Z = 0.001$ than with $Z = 0.008$, during the time $\mathscr{L}_{mech}(t)$ is dominated by stellar winds. This occurs because stellar wind strength depends strongly on metallicity. The mechanical luminosity $\mathscr{L}_{mech}(t)$ with $Z = 0.001$ soon approaches $\mathscr{L}_{mech}(t)$ expected with $Z = 0.008$. In this study, we are specifically interested in the evolution of bubbles at early time before they blow out of the disks, at $t \lesssim 10$ Myr at $z = 0$ and $t \lesssim 3$ Myr at $z = 3$ and 8 (see § 5.2 for discussion on bubble dynamics).

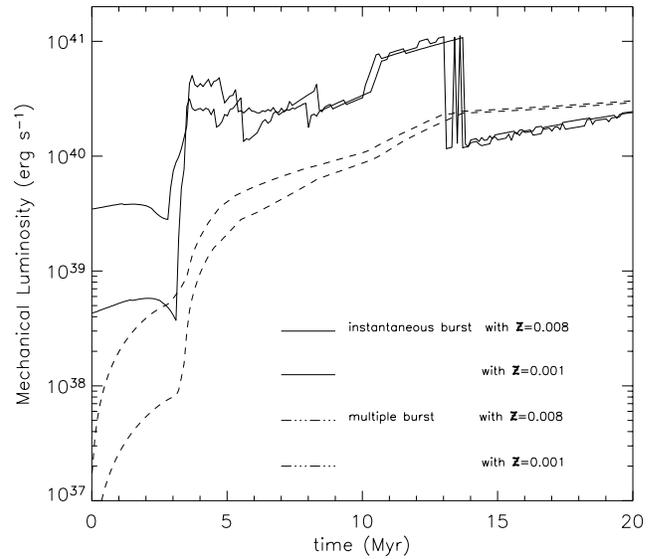

Fig. 6.—Mechanical luminosities $\mathscr{L}_{mech}(t)$ as a function of time, drawn from the Starburst 99 model (Leitherer et al. 1999) when $10^6\ M_\odot$ of gas is converted to stars, for the instantaneous-burst model with metallicities $Z = 0.008$ (*thick solid line*) and $Z = 0.001$ (*thin solid line*) and for the multiple-burst model with $Z = 0.008$ (*thick dashed line*) and $Z = 0.001$ (*thin dashed line*).

The average mechanical luminosity $L_{mech}$ within 10 Myr after the onset of starburst with $M_{SB} = 10^6\ M_\odot$ is $L_{mech}(10^6\ M_\odot) \approx 2 \times 10^{40}$ ergs s$^{-1}$ for the instantaneous-burst model with $Z = 0.008$ (see Fig. 6, *thick solid line*). This is used as a reference value for the other three cases. We lower $L_{mech}$ by 1 order of magnitude from the reference value if the multiple-burst model is applied with the same metallicity and if the instantaneous-burst model is applied with $Z = 0.001$. We lower $L_{mech}$ by 2 orders of magnitude from the reference value if the multiple-burst model is applied with $Z = 0.001$. Note the weak dependence of bubble radius on mechanical luminosity in a uniform medium; $R_{bl} \propto L_{mech}^{1/5}$ (Castor, McCray, & Weaver 1975). We vary the mechanical luminosity linearly with the total mass of gas converted to stars, $M_{SB}$: $L_{mech}(M_{SB}) = L_{mech}(10^6\ M_\odot) M_{SB}/(\xi \times 10^6\ M_\odot)$ with $\xi = 2$.

We model the effects of starbursts on dwarf galactic disks with $M_d = 10^8$, $10^9$, and $10^{10}$ $M_\odot$ at $z = 0$ and with $M_d = 10^8$ $M_\odot$ at $z = 3$ and 8. Because of limited computational resources, we simulate the evolution of bubbles only until they blow out of the disks and create galactic outflows. Our aim is (1) to test the effects of supershells on the entrapment of photons while the shells are within the disks, that is, before the blowouts, and (2) to test whether galactic outflows increase the escape of photons after the blowouts. The opening angles of chimneys created by the outflows are assumed to stay constant after they blow out.

## 5. RESULTS

### 5.1. Undisturbed Disks

We first analyze the results for disks without outflows. We define the total escape fraction of photons from a disk as

$$F_{\rm esc} = \lim_{t\to\infty} F_{\rm esc}(t) = \lim_{t\to\infty} \frac{\int_0^t f_{\rm esc}(t) N_{\rm ph}(t) dt}{\int_0^t N_{\rm ph}(t) dt}, \quad (16)$$

where $N_{\rm ph}(t)$ is a given photon luminosity, with the instantaneous escape fraction

$$f_{\rm esc}(t) = \frac{1}{N_{\rm ph}(t)} \left[ N_{\rm ph}(t) - \int_{-\pi/2}^{\pi/2} d\sin\theta\, d\theta \int_0^{2\pi} d\varphi \right.$$
$$\left. \times \int_0^{r_{\rm I}(\theta,\varphi)} \alpha_B n_p^2(r, \theta, \varphi) r^2\, dr \right]. \quad (17)$$

The last integration gives the total effective recombination rate, where $r_{\rm I}$ is the radius of the ionization front at time $t$. Neither the total nor the instantaneous escape fractions depend on the viewing angle of the observer. It is important to remember that observers measure the instantaneous escape fraction $f_{\rm esc}(t)$, not the total escape fraction $F_{\rm esc}$. We show later in this section that $f_{\rm esc}(t)$ and $F_{\rm esc}$ can differ drastically when the Lyman continuum spectrum depends strongly on time.

Figure 7 shows the fraction of photons escaping from disks as a function of star formation efficiency $f_*$ with our standard model. With the maximum star formation efficiency that we consider, $f_* = 0.6$, which roughly corresponds to the critical mass set by the $10^{21}$ cm$^{-2}$ cutoff, most of the photons ($\gtrsim 50\%$) can escape in all the disks at all redshifts with any given spectrum of Lyman continuum radiation, except in thick disks at $z = 8$. The total escape fractions $f_{\rm esc}$ decrease as a decreasing function of $f_*$ but vary with the different assumptions for disk morphology and star formation history that we apply. We discuss below the dependence of various parameters in our study on the escape of ionizing photons.

### 5.1.1. Scale Heights

The thick disks can trap the ionizing radiation more effectively in many cases, especially at high redshift, while thin disks let more radiation out. This is because the photons find their way out of the disks in the vertical direction first, while they are often trapped in the radial direction ($\Xi_Z < \Xi_R$). The effective recombination rate in the vertical direction is $\Xi_Z \propto H$, from equation (3). Recall that the presence of disk potentials makes the disks unreasonably thin at high redshift; therefore, the results with thin disks should be ignored at $z \gg 0$ but can still be compared with the results from thick disks. There are some cases in which the thick disks yield slightly higher $f_{esc}$ than the thin disks, when the ionization front immediately propagates through the entire disks, while $\Xi_R > \Xi_Z$ and $\Xi_R \propto H^{-2}$ from equation (4).

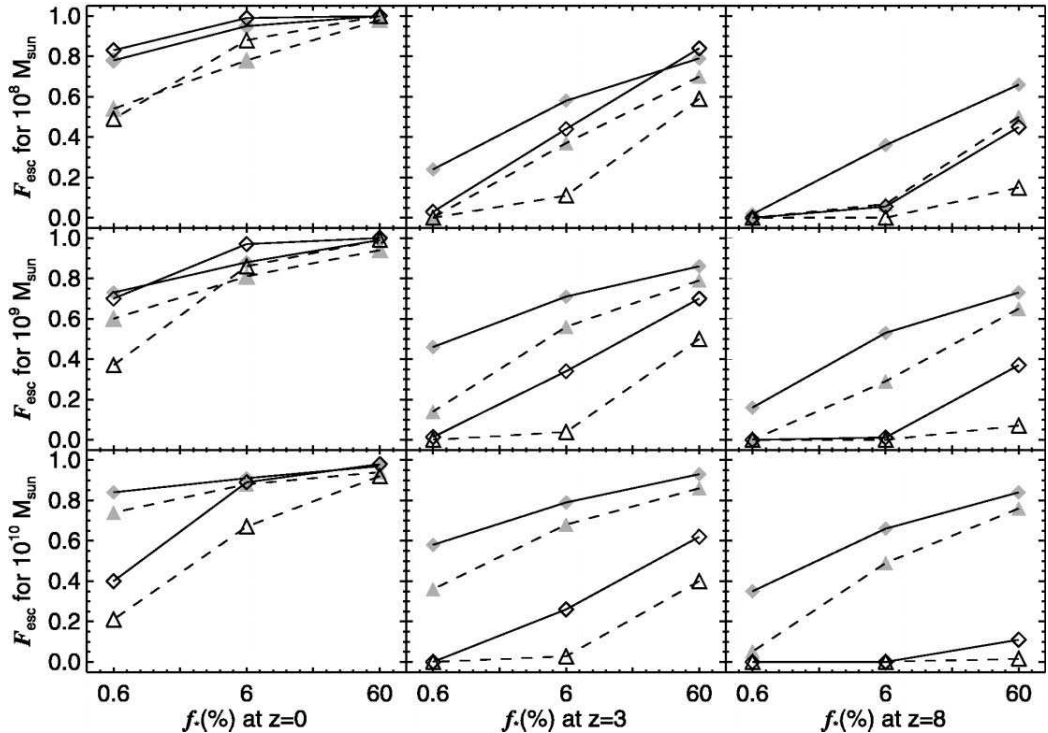

FIG. 7.—Total escape fractions $F_{\rm esc}$ for models without bubbles with thin (*filled symbols*) and thick (*open symbols*) disks and with the instantaneous-burst model (*diamonds and solid lines*) and the multiple-burst model (*triangles and dashed lines*). The total escape fractions $F_{\rm esc}$ decrease as redshift increases and do not depend much on the star formation model applied. Thick disks can trap more ionizing photons than thin disks in general.

### 5.1.2. Redshift and Mass

The escape fractions decrease as redshift increases in undisturbed disks (see Fig. 7), confirming previous work by Wood & Loeb (2000) and Ricotti & Shull (2000). The trend is due to the increase of the surface density, $\Sigma \propto R_d^{-2} \propto (1+z)^2$. Recall $\Xi_Z \propto R_d^{-4} H \propto (1+z)^4 H$, so the effect of an increase in the redshift is substantially greater than that of a decrease in $H$ (see Table 1). However, we show in § 5.2 that the redshift dependence for a given $f_*$ no longer exists once the effects of galactic outflows are taken into account. On the other hand, we do not see the trend of decreasing escape fractions as the disk masses increase, as in Wood & Loeb (2000). The escape fractions at a given redshift are approximately the same for all our disks with a given $f_*$, because $\Xi_Z \propto \Sigma_0^2 H \propto M^{2/3} H$ and $N_{ph} \propto f_* M$.

### 5.1.3. Star Formation History

There is a mild increase in the escape fractions from the multiple-burst model to the instantaneous-burst model. The trend reflects the behaviors of photon luminosities $N_{ph}(t)$, which are an order of magnitude higher during the first few megayears with the instantaneous-burst model than with the multiple-burst model (see Fig. 1). We show in Figure 8 the instantaneous escape fraction $f_{esc}(t)$ and the cumulative total escape fraction $F_{esc}(t)$ as a function of time in a disk with $M_d = 10^8\ M_\odot$ at $z = 0$ with $f_* = 0.006$ and $0.06$ and two different models of Lyman continuum history. The shape of $f_{esc}(t)$ as a function of time reflects that of $N_{ph}(t)$.

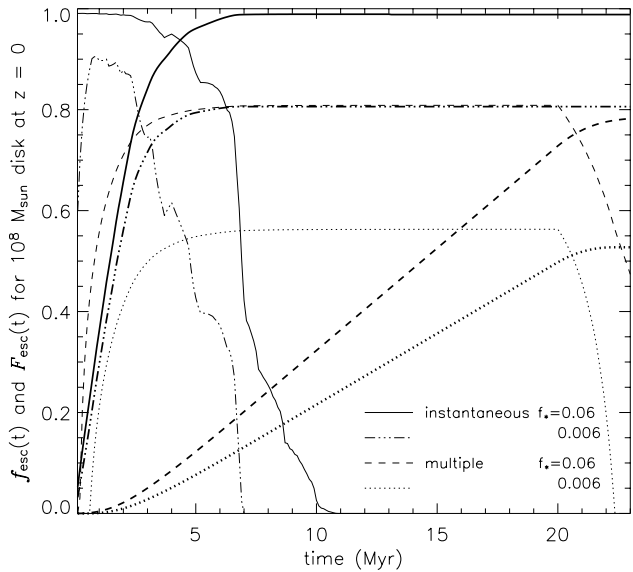

Fig. 8.—Instantaneous escape fractions $f_{esc}(t)$ (*thin lines*) and the total cumulative escape fractions $F_{esc}(t)$ (*thick lines*) plotted for a model with $M_d = 10^8\ M_\odot$ at $z = 0$ with Lyman continuum histories, based on the instantaneous-burst model and the multiple-burst model. The star formation efficiencies $f_*$ applied are 0.06 (*solid and dashed lines*) and 0.006 (*triple-dot–dashed line and dotted line*). The figure shows that the observed value of the escape fraction at time $t$, $f_{esc}(t)$, does not correspond to the total escape fraction $F_{esc} = \lim_{t\to\infty} F_{esc}(t)$.

For example, $f_{esc}(t)$ peaks sharply in the instantaneous-burst model within the first few megayears, as $N_{ph}(t)$ does, and as a result most Lyman continuum photons escape the disk quickly, yielding the total escape fraction $F_{esc} \gtrsim 0.8$. On the other hand, $f_{esc}(t)$ rises slowly to its peak in the multiple-burst model, yielding $F_{esc} \approx f_{esc}(t) < 0.8$. We emphasize that the instantaneous escape fraction $f_{esc}(t)$ that observers can measure does not necessarily reflect the total fraction of ionizing photons that escape out of starbursting galaxies $F_{esc}$ if the bursts are nearly instantaneous.

$F_{esc}$ and $f_{esc}(t)$ with a different star formation history can be inferred based on the results above. The peak photon luminosity primarily determines the escape fraction from a starburst clump. Therefore, $F_{esc}$ and $f_{esc}(t)$ from multiple starbursts with a longer duration or a constantly star-forming clump with a correspondingly lower peak photon luminosity are expected to be smaller. However, they do not scale linearly with the peak photon luminosity, because of disk geometry.

### 5.1.4. Distribution of Ionizing Sources

Recall that our study is limited to single starburst clusters or ionizing sources at the center of our model disks. These central clusters convert $f_* M_d$ of gas into stars, and so $f_*$ in our study does not correspond to star formation efficiencies of the entire disks. We also argued in § 3.2 that we compute lower limits to the escape fractions in our study by placing the starburst clusters at the center and that the escape fractions remain as lower limits when there are multiple starburst clusters of the same luminosities scattered around the disks. Below, we give some insights into a situation in which a given amount of gas is available for star formation throughout a given disk.

Dove & Shull (1994) and Dove et al. (2000) modeled the Galactic disk as having randomly distributed OB associations. The total escape fraction from the Galactic disk is found to be ~0.1. In the study by Wood & Loeb (2000), starburst ionizing sources are spread throughout the disks with emissivity scaling as $n_{HI}^2$ (like OB associations) or $n_{HI}$, and quasar sources are located at the center of the disks. At $z > 5$, more than 0.3 of Lyman continuum photons escape from their miniquasars and none escape from their starburst galaxies. The low escape fractions seen in some of their model disks are due to the distribution of ionizing sources.

In a given disk, there is a minimum photon luminosity $N_{ph,min}$ for any photon to escape or for a Strömgren sphere to break out of it, as is also mentioned in Dove & Shull (1994). This means when a fixed amount of mass is converted to stars in a galaxy, having numerous or multiple ionizing sources is not a favorable condition for the photons to escape, since many OB associations then have photon luminosities below $N_{ph,min}$. In Table 3, we list $N_{ph,min}$ for our model disks. Note that $N_{ph,min}$ is equal to the minimum effective recombination rate, that is, $\Xi_Z \propto N_{HI}^2 H$, times $4\pi$. Therefore, $N_{ph,min}$ is higher for more massive disks and disks at high redshift and is also higher for thicker disks. We conclude that when a fixed amount of mass is available for starburst, placing it all in a single source gives the most favorable condition for the escape of ionizing radiation.

### 5.1.5. Other Issues

Changing the cosmology from an SCDM model ($\Omega_0 = 1$ and $h = 0.5$) to a $\Lambda$CDM model ($\Omega_0 = 0.37$ and $h = 0.7$)

TABLE 3
Minimum Photon Luminosity for a Photon to
Escape from Model Disk

| $M_d$ ($M_\odot$) | Thin Disk, $z = 0$ ($10^{52}$ s$^{-1}$) | Thick Disk, $z = 0$ ($10^{52}$ s$^{-1}$) | $z = 3$ ($10^{52}$ s$^{-1}$) | $z = 8$ ($10^{52}$ s$^{-1}$) |
|---|---|---|---|---|
| $10^8$ | 0.040 | 0.16 | 1.3 | 9.6 |
| $10^9$ | 0.059 | 0.58 | 19 | 140 |
| $10^{10}$ | 0.022 | 0.87 | 280 | 2100 |

Note.—The minimum photon luminosity for any photon to escape from the model disk is $N_{\mathrm{ph,min}} = 4\pi\Xi_z$.

changes the virial radius of the halos, $R_{\mathrm{vir}} \propto [h^2\Omega_0/\Omega(t)]^{-1/3}$, and therefore the scale radius $R_d \propto R_{\mathrm{vir}}$ and the central surface densities of the disks $\Sigma_0 \propto R_d^{-2}$. At high redshift $\Omega(t) \sim 1$, so the changes in $h^2$ and $\Omega_0$ cancel out. Any significant difference between cosmologies appears only for low-redshift disks at $z \ll 1$ with $\Omega(t) = \Omega_0$. Even there, the maximum increase in $\Sigma_0$ is only a factor of 2. We find in our low-redshift $\Lambda$CDM models that the qualitative results on the total escape fractions $F_{\mathrm{esc}}$ do not change.

We also checked the effect of increasing the disk gas in low-mass halos at all redshifts by using the cosmological dark-to-baryonic ratio $\phi = 7.4$ instead of $\phi_{\mathrm{lcl}}$. A high value of $\phi$ means a decrease in halo mass for a given disk mass. The virial radius and the scale radius of the new halo decrease ($R_d \propto M_h^{-1/3}$), and the surface density of the disk increases up to a factor of a few. We find that there is no significant change in $F_{\mathrm{esc}}$ from our model disks with $\phi = 7.4$.

The surface density increases only by a factor of a few when the spin parameter of a halo differs by 1 $\sigma$ from the mean: $\Sigma_0 \propto \lambda^{-2} = (0.035/0.05)^{-2} \approx 2$. The above results with different cosmology and values of $\phi$ suggest that the change in the angular momentum of a disk within a 1 $\sigma$ range will not influence the qualitative trends of $F_{\mathrm{esc}}$ seen in Figure 7. Any change in $\lambda$ much greater than 1 $\sigma$ should be compensated for by the change in star formation efficiency $f_*$, as we discussed in § 3.2.

If we use the dark matter profile by Navarro, Frenk, & White (1997) with the halo concentration factor, $c = 5$–30 (Mo 1998), instead of the dark matter profile in equation (6), the maximum decrease in scale radius $R_d$ is expected to be less than $\frac{1}{2}$. The increase in $\Xi_Z$ is then limited to less than 1 order of magnitude.

### 5.1.6. Summary

We can draw our results from Figure 7 together by considering the effective recombination rate in the vertical direction from equation (3): $\Xi_Z \propto R_d^{-4} H \propto (1+z)^4 H$. The total escape fraction of ionizing radiation $F_{\mathrm{esc}}$ decreases as redshift $z$ increases. For a disk with given mass $M_d$ and column density distribution $\Sigma(R)$ at a given redshift, $F_{\mathrm{esc}}$ also decreases as the scale height $H$ of the disk increases. Changes in scale radius $R_d$ due to changes in cosmology, dark-to-baryonic matter ratio, spin parameter $\lambda$ by $\sim$1 $\sigma$, or dark matter distribution produce only small changes in $F_{\mathrm{esc}}$. As we argued in § 3.2, a further increase (decrease) in $R_d$, for example, with a greater than 1 $\sigma$ increase (decrease) in $\lambda$ should be compensated with a decrease (increase) in $f_*$ and $N_{\mathrm{ph}}$. Therefore, we conclude that the escape of ionizing radiation is sensitive to vertical gas structures rather than radial gas structures for fixed $M_d$, specifically the scale height $H$.

Despite the sensitivity of the escape fraction to disk morphology, the results converge at low redshift. At $z = 0$, most of the photons easily escape even with star formation efficiency $f_*$ as low as $\sim$0.006. Recall we are solving for the lower limits on the escape fractions for starburst clusters. Our results strongly suggest that most photons do escape from dwarf galactic disks at low redshift if bubble dynamics are not considered. These results seem to contradict the observations of local dwarf starburst galaxies with very low escape fractions, $f_{\mathrm{esc}}(t) < 0.06$. However, we show in the next section that the formation of shells by starburst bubbles significantly alters the results.

### 5.2. Disks with Shells and Galactic Outflows

We model the effects of stellar winds and SNe from starbursts in our dwarf galactic disks to examine how swept-up shells and galactic outflows affect the escape of ionizing radiation. We experiment with disks with $M_d = 10^8$–$10^{10} M_\odot$ at $z = 0$ and $M_d = 10^8 M_\odot$ at $z = 3$ and 8. We show below two significant results of our experiments: (1) swept-up shells of neutral hydrogen can substantially trap the ionizing radiation before blowouts, even in disks at low $z$ that otherwise yield high $f_{\mathrm{esc}}(t)$ and $F_{\mathrm{esc}}$ and (2) galactic outflows allow a significant fraction of ionizing radiation to escape directly to the IGM from disks at high redshift that otherwise yield $F_{\mathrm{esc}} \approx 0$.

#### 5.2.1. Low-Redshift Thin Disks

Figure 9 shows the propagation of ionization fronts overlaid on density distributions in a thin disk with $M_d = 10^8 M_\odot$ at $z = 0$, with $f_* = 0.06$ by using the multiple-burst model (with $\mathscr{Z} = 0.008$) at $t = 2$ and 6 Myr after the onset of the starburst. We take $L_{\mathrm{mech}} = 10^{40}$ ergs s$^{-1}$, 1 order of magnitude lower than the mechanical luminosity expected with the instantaneous-burst model (see Fig. 6), to examine the effects of shells swept up by a bubble before it begins to blow out ($t \lesssim 5$ Myr).

The swept-up shell has density $n_d(R, Z)\sqrt{2}\mathscr{M}_A < n_{\mathrm{sh}} < n_d(R, Z)\mathscr{M}^2$ because of the isothermal shock. As we mentioned in § 4, we compensate for the inability to fully resolve the density jump in the shells in our simulations by computing the effective recombination rate, $\Xi_{\rm sh} \propto n_{\rm sh} N_{\rm sh} r_{\rm sh}^2$, at each angle through the shells in our photoionization code, using the hydrodynamical value for $n_{\rm sh}$ to maintain a lower limit. At $t = 2$ Myr in Figure 9 (*left*), the dense shell completely traps the ionizing radiation, so $f_{\rm esc}(t) = 0$. By $t = 6$ Myr (*right*), the bubble starts to accelerate in the vertical direction, where the disk gas is most stratified, so the Rayleigh-Taylor instability begins to fragment the shells. The fragmentation opens funnels for the photons to escape, so $f_{\rm esc}(t) = 0.31$. By that time, the accelerated shock in the vertical direction is no longer isothermal. The further evolution of the bubble leads only to an increase in $f_{\rm esc}(t)$, because the bubble eventually blows out with more fragmentation and the mechanical luminosity of the burst is expected to increase up to $10^{41}$ ergs s$^{-1}$ after $t \sim 10$ Myr.

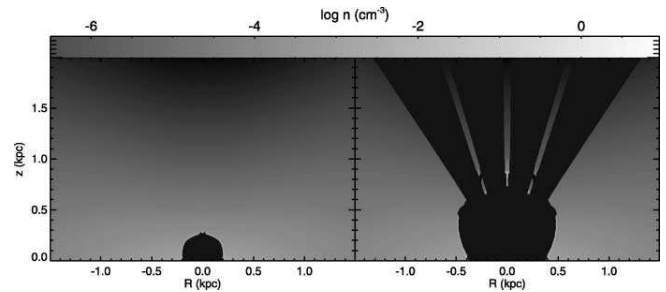

Fig. 9.—Entrapment and the escape of ionizing radiation shown in black on top of density distributions in a thin disk with $M_d = 10^8 M_\odot$ at $z = 0$, at $t = 2$ and 6 Myr after the onset of starburst with $f_* = 0.06$. The Lyman continuum history applied is the multiple-burst model with $\mathscr{Z} = 0.008$, which corresponds to an average mechanical luminosity $L_{\mathrm{mech}} = 10^{40}$ ergs s$^{-1}$ within 5–10 Myr. The cold dense shells of swept-up ISM completely trap the ionizing photons at $t = 2$ Myr (left) but begin to fragment as the bubble accelerates by $t = 6$ Myr (right). The fragmentation creates funnels for the photons to escape to the IGM.

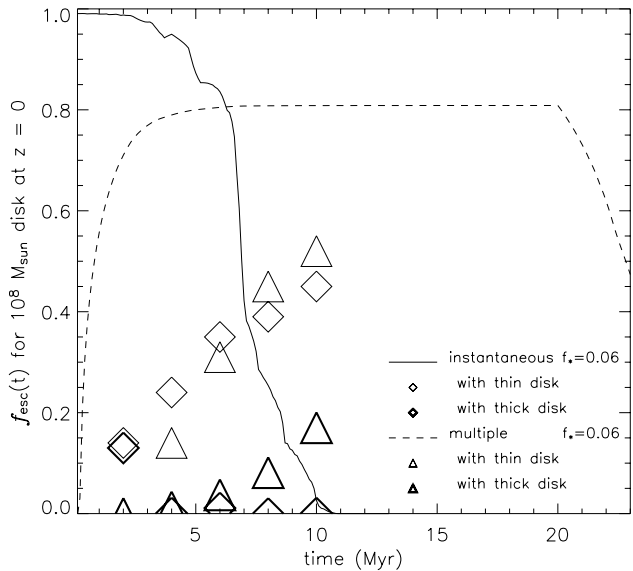

Fig. 10.—Changes in the instantaneous escape fractions $f_{\rm esc}(t)$ when the effect of shells of swept-up ISM is included. The model is a disk with $M_d = 10^8\ M_\odot$ at $z = 0$, with $f_* = 0.06$, and the Lyman continuum photon luminosities are based on the instantaneous-burst model (*diamonds and solid line*) with $L_{\rm mech} = 10^{41}$ ergs s$^{-1}$ and the multiple-burst model (*triangles and dashed line*) with $L_{\rm mech} = 10^{40}$ ergs s$^{-1}$. The instantaneous escape fractions $f_{\rm esc}(t)$ with shells cooled to the density expected for isothermal shocks are shown with open symbols, $f_{\rm esc}(t)$ with unresolved shells in our lower resolution simulations are shown with filled symbols, and $f_{\rm esc}(t)$ without shells are shown with lines. The decrease in $f_{\rm esc}(t)$ is significant once the maximum effect of fully cooled shells is included, before the bubble blows out of the disk at $t \approx 8$ Myr.

Figure 10 shows the instantaneous escape fraction $f_{\rm esc}(t)$ for the model shown in Figure 9 with $f_* = 0.06$ with shells cooled to the theoretical density $n_{\rm sh} = n_d \mathcal{M}^2$ (*open symbols*), with the unresolved shells from our simulations (*filled symbols*), and without shells (*lines*). We compute $f_{\rm esc}(t)$ every 2 Myr up to 10 Myr in our simulation with a resolution of 2.5 pc. The photon luminosity of the instantaneous-burst model reaches its peak, $\sim 2 \times 10^{53}$ s$^{-1}$, already at $t = 0.2$–0.3 Myr and decreases by more than an order of magnitude by $t \sim 5$ Myr. That of the multiple-burst model is $\sim 2 \times 10^{52}$ s$^{-1}$ by $t \sim 2$ Myr, still rising very slowly up to $\sim 3 \times 10^{52}$ s$^{-1}$ for the duration of starburst ($\tau_{\rm SB} = 20$ Myr).

A significant effect from the underdensity of the shell due to numerical resolution appears only at $t \lesssim 4$ Myr, before the bubble begins to blow out of the disk. The shells can trap the photons very effectively while the bubble is within a few scale heights of the disk ($H = 100$ pc). Afterward the shells fragment as the bubble accelerates, allowing the photons to escape. Figure 10 shows that the proper treatment of shell formation is crucial to study the full effect of the shells on the escape of ionizing photons.

With the fully cooled shells, we significantly suppress $f_{\rm esc}(t)$. As a result, the total escape fractions $F_{\rm esc}$ decrease from 0.95 to 0.16 with the instantaneous-burst model and from 0.78 to 0.42 with the multiple-burst model. We assume that $f_{\rm esc}(t)$ stays constant after $t \geq 10$ Myr because the opening angle for the escape of photons is primarily determined by that of the bubble after blowout, which remains fairly constant at $\sim 30°$–$40°$. The value of $F_{\rm esc}$ computed with the multiple-burst model is strictly a lower limit, since the mechanical output of the burst increases by an order of magnitude at later time (see Fig. 6). However, recall that the bubble evolution depends only weakly on $L_{\rm mech}$: $R_{\rm bl} \propto L_{\rm mech}^{1/5}$. The decrease in $F_{\rm esc}$ is more significant with the instantaneous-burst model than the multiple-burst model. This is because the shells significantly trap ionizing photons only while the bubble is still within a few scale heights, and most photons are produced during that time with the instantaneous-burst model.

### 5.2.2. Low-Redshift Thick Disks

Bubble evolution depends sensitively on the scale height $H$ in our model disks. The results for $f_{\rm esc}(t)$ and $F_{\rm esc}$ are different in a thick disk with a larger $H$. As mentioned above, the shells of swept-up ISM are very effective in trapping radiation before acceleration starts above a few scale heights. The time when the bubble reaches $H$ in the vertical direction can be approximated as $t_H \propto L_{\rm mech}^{-1/3} n_0^{1/3} H^{5/3}$ in a uniform medium with a constant density, $n_0$ (Castor et al. 1975). We can approximate $n_0 = n_* \approx N_{\rm H\,I,0}/H$, that is, the central density of the Gaussian or the exponential profile. Then for constant column density $N_{\rm H\,I,0}$, $t_H \propto H^{4/3}$. The scale height of the thin disk with dark matter and disk potentials is 100 pc, but that of the thick disk with only dark matter potential is 580 pc, both with $M_d = 10^8\ M_\odot$ at $z = 0$. Therefore, the time for the bubble to reach $H$ is a factor of 10 shorter in the thin disk than in the thick disk. In addition, the vertical gas distribution is exponential in the thin disk because of the additional gravitational pull by the disk, while it is Gaussian in the thick disk. The stratification drives the acceleration of a bubble (Kompaneets 1960; Mac Low & McCray 1988), so the bubble blows out of the thin disk much earlier than the thick disk. On the other hand, the shock velocity $v_{\rm sh}$ at $H$ is proportional to $L_{\rm mech}^{1/3} H^{-1/3}$, so the Mach number is a factor of 2 larger in the thin disk than in the thick disk. Recall that $\Xi_{\rm sh} \propto n_{\rm sh}$ and $n_{\rm sh} = n_d \mathcal{M}^2$, but this decrease in $\Xi_{\rm sh}$ is small compared with the increase in the time for a blowout in the thick disk. The opening angles of the chimneys cleared out by Rayleigh-Taylor instability is $\sim 30°$–$40°$ at the time of blowout in both disks.

Figure 11 shows that the instantaneous escape fractions $f_{\rm esc}(t)$ are significantly suppressed in the thick disk with a larger scale height, until the bubble begins to accelerate at $t \approx 10$ Myr. The instantaneous escape fractions $f_{\rm esc}(t)$ are sensitive to the different scale heights, $H$, because blowouts determine the transition from the entrapment of photons by shells to the escape of photons through chimneys. The computed total escape fractions $F_{\rm esc}$ are 0.16 in the thin disk and 0.11 in the thick disk with the instantaneous-burst model and 0.42 and 0.12, respectively, with the multiple-burst model.

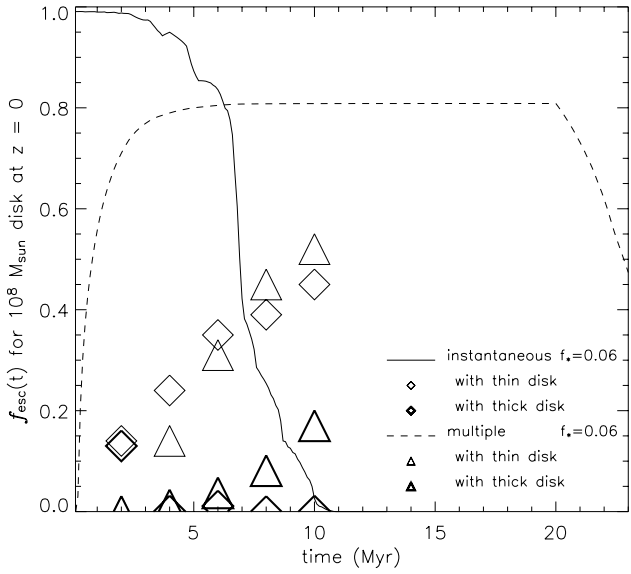

Fig. 11.—Instantaneous escape fractions $f_{esc}(t)$ in the thick disk with $H = 580$ pc (*bold diamonds and bold triangles*), compared with those in the thin disk with $H = 100$ pc (*diamonds and triangles*) for the same galaxy shown in Fig. 10. The thick disk traps more photons than the thin disk until $t \approx 10$ Myr, because the growth of the bubble in the thick disk is slower than that in the thin disk.

Experiments with high-mass disks with $M_d = 10^9$ and $10^{10}$ $M_\odot$ at $z = 0$ with $f_* = 0.06$ show that the qualitative results are the same: $f_{esc}(t)$ is suppressed only before blow-out. The scale height of the disk strongly determines the time of blowout. The scale heights of the thin disks are 56 and 19 pc and those of the thick disks are 750 and 980 pc with $M_d = 10^9$ and $10^{10}$ $M_\odot$, respectively. Therefore, the bubbles in thick disks can trap the radiation up to $t \sim 10$ Myr after the onset of starbursts, but the bubbles in thin disks immediately blow out of the disks and create chimneys for the escape of photons. As a result, the total escape fractions $F_{esc}$ are significantly more suppressed in thick disks with large $H$ than in the thin disks with small $H$ (see Table 4)

### 5.2.3. High-Redshift Disks

At high redshift, halos are very dense [$\rho_h \propto (1+z)^3$] and disks are compact [$R_d \propto (v)^{-1}$ and $\Sigma_0 \propto (1+z)^2$]; therefore, the strong gravitational field of the halos creates highly stratified atmosphere in the disks even without disk potentials (thick disks in our study). As a result, bubbles quickly blow out of the disks at very early time, about a few megayears after the onset of starbursts, although the mechanical power of the bursts is expected to be smaller, while stellar winds dominate in high-redshift disks with $\mathscr{Z} \approx 0.001$ than in low-redshift disks with $\mathscr{Z} \approx 0.008$.

Figure 12 shows the evolution of a starburst bubble and the propagation of ionization fronts at $t = 2$ and 3 Myr in a disk with $M_d = 10^8$ $M_\odot$ at $z = 8$, with $f_* = 0.06$ and the multiple-burst model. We expect lower metallicity at high redshift, $\mathscr{Z} \lesssim 0.001$; therefore, the star formation efficiency $f_* = 0.06$ with the multiple-burst model corresponds to $L_{mech} = 10^{39}$ ergs s$^{-1}$ within the first few Myr. This is because the stellar winds dominate the mechanical luminosity during the time, and the production of stellar winds is very sensitive to the metallicity (see Fig. 6). We apply $L_{mech} = 10^{39}$ ergs s$^{-1}$ because we are specifically interested in the

TABLE 4
Total Escape Fractions

| $M_d$ ($M_\odot$) | $z$ | $f_*$ | Model | Instantaneous Burst $L_{mech}$ (ergs s$^{-1}$) | Shell | None | Multiple Burst $L_{mech}$ (ergs s$^{-1}$) | Shell | None |
|---|---|---|---|---|---|---|---|---|---|
| $10^8$ | 0 | 0.06 | Thin | $10^{41}$ | 0.16 | 0.95 | $10^{40}$ | 0.42 | 0.78 |
| | | | Thick | $10^{41}$ | 0.11 | 0.99 | $10^{40}$ | 0.12 | 0.88 |
| | | 0.006 | Thin | $10^{40}$ | 0.081 | 0.78 | $10^{39}$ | 0.17 | 0.54 |
| | | | Thick | $10^{40}$ | 3.5E−5 | 0.83 | $10^{39}$ | 0.033 | 0.49 |
| $10^9$ | 0 | 0.06 | Thin | $10^{42}$ | 0.54 | 0.88 | $10^{41}$ | 0.63 | 0.81 |
| | | | Thick | $10^{42}$ | 0.017 | 0.97 | $10^{41}$ | 0.29 | 0.86 |
| | | 0.006 | Thin | $10^{41}$ | 0.37 | 0.73 | $10^{40}$ | 0.53 | 0.60 |
| | | | Thick | $10^{41}$ | 0.022 | 0.70 | $10^{40}$ | 0.017 | 0.37 |
| $10^{10}$ | 0 | 0.06 | Thin | $10^{43}$ | 0.87 | 0.91 | $10^{42}$ | 0.84 | 0.88 |
| | | | Thick | $10^{43}$ | 0.0097 | 0.89 | $10^{42}$ | 0.21 | 0.67 |
| | | 0.006 | Thin | $10^{42}$ | 0.85 | 0.84 | $10^{41}$ | 0.80 | 0.74 |
| | | | Thick | $10^{42}$ | 0.0004 | 0.40 | $10^{41}$ | 0.081 | 0.21 |
| $10^8$ | 3 | 0.6 | | $10^{41}$ | 0.49 | 0.84 | $10^{40}$ | 0.56 | 0.59 |
| | | 0.06 | ... | $10^{40}$ | 0.13 | 0.44 | $10^{39}$ | 0.20 | 0.11 |
| | | 0.6 | | $10^{39}$ | 0.0029 | 0.030 | $10^{38}$ | 0.21 | 0.0 |
| $10^8$ | 8 | 0.6 | | $10^{41}$ | 0.62 | 0.45 | $10^{40}$ | 0.50 | 0.15 |
| | | 0.06 | ... | $10^{40}$ | 0.28 | 0.056 | $10^{39}$ | 0.22 | 0.0 |
| | | 0.006 | | $10^{39}$ | 0.07 | 0.0 | $10^{38}$ | 0.0 | 0.0 |

Note.—Total escape fractions $F_{esc}$ are computed in the disks with supershells and galactic outflows created by starburst bubbles (shell), compared with $F_{esc}$ in undisturbed disks (none). Only thick disks are considered at $z > 0$. The metallicities assumed are $\mathscr{Z} = 0.008$ at $z = 0$ and $\mathscr{Z} = 0.001$ at $z > 0$.

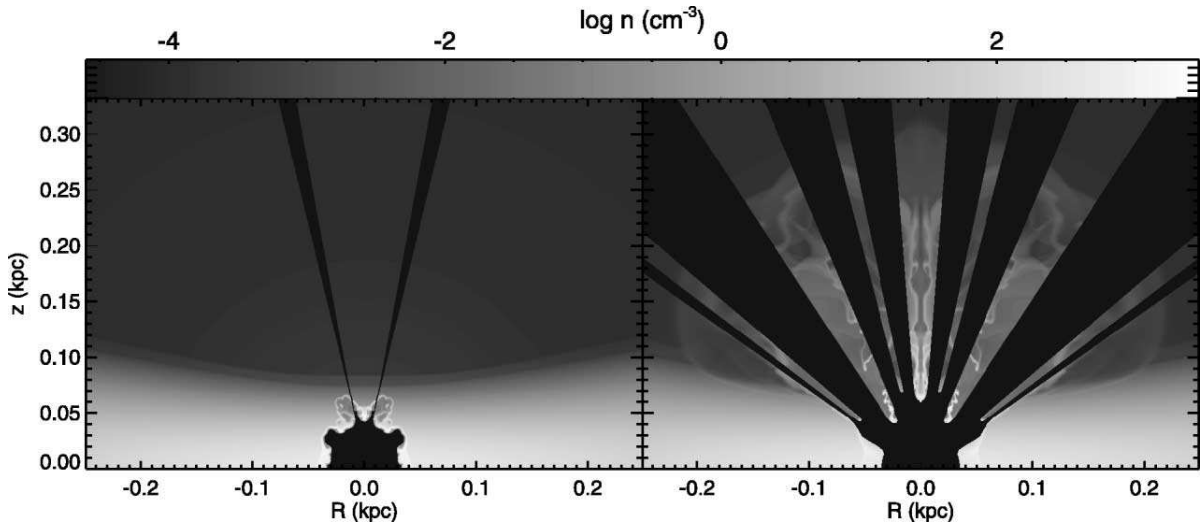

Fig. 12.—Starburst bubble in a disk with $M = 10^8\ M_\odot$ at $z = 8$, with $f_* = 0.06$, the multiple-burst model, and a corresponding mechanical luminosity of $10^{40}$ ergs s$^{-1}$. The bubble sweeps the high-redshift dense ISM into very dense shells up to $t = 2$ Myr (left), which are very effective in trapping the ionizing radiation. The bubble, however, immediately blows out of the disk by $t = 3$ Myr (right). The chimneys created by the outflow let $\sim$20% of ionizing photons produced by stars escape to the IGM.

evolution of bubbles at early time before they blow out of the disks and the shells fragment. Recall that we experiment with thick disks only at high redshift. Figure 12 shows that the bubble immediately breaks out of the disk by $t = 3$ Myr, even with the very small mechanical power supplied by low-metallicity stellar winds, and lets the photons escape through the chimney created, so $f_{\rm esc}(t) = 0.014$ at $t = 2$ Myr, but $f_{\rm esc}(t) = 0.24$ at $t = 3$ Myr. The opening angles of the chimneys cleared out by Rayleigh-Taylor instability are $\sim$40°–50° at $t = 3$–5 Myr. The fragmented shells are already falling back and prevent the escape of photons; however, after $t = 3$ Myr, more and more supernovae go off, supplying more mechanical power for the bubble to grow (see Fig. 6), so we expect the escape fraction to remain $f_{\rm esc}(t) \gtrsim 0.2$. Remember that only a negligible fraction of photons can escape out of undisturbed high-redshift disks because of their high density.

Figure 13 shows the increase in $f_{\rm esc}(t)$ when the effects of galactic outflows are taken into account in disks with $M_d = 10^8\ M_\odot$ at $z = 3$ and at $z = 8$ with $f_* = 0.06$ and $L_{\rm mech} = 10^{40}$ ergs s$^{-1}$ for the instantaneous-burst model and $L_{\rm mech} = 10^{39}$ ergs s$^{-1}$ for the multiple-burst model. The peak luminosities are the same as before. At $z = 8$, the escape fractions $f_{\rm esc}(t)$ are nearly zero with both Lyman continuum histories in undisturbed disks, but $f_{\rm esc}(t)$ increases to greater than 0.2 as the bubbles blow out of the disk at a very early time, by $t \sim 1$ Myr (instantaneous) and $\sim$3 Myr (multiple). At $z = 3$, $f_{\rm esc}(t)$ are rather high with the instantaneous-burst model but are less than 0.15 with the multiple-burst model in undisturbed disks, but $f_{\rm esc}(t)$ are greater than 0.2 as the bubbles blow out by $t \sim 3$ and $\sim$6 Myr, respectively. We did not further compute the evolution of the bubbles after they blow out, because of the great computational expense. However, we do not need to compute $f_{\rm esc}(t)$ after blowouts if the starbursts are instantaneous. When the starbursts last for 20 Myr, we assumed that the sizes of the chimneys for the escape of photons do not change much after blowouts.

We find that early blowouts in dense, high-redshift disks create chimneys for the photons to escape freely to the IGM. As a result, the total escape fractions $F_{\rm esc}$ are 0.28 (0.056 in the same but undisturbed disk) with the instantaneous-burst model and 0.22 (0.0) with the multiple-burst model in the disk at $z = 8$, and are 0.13 (0.44) and 0.20 (0.11), respectively, in the disk at $z = 3$. The total escape fraction $F_{\rm esc}$ is suppressed with the instantaneous-burst model at $z = 3$ because most ionizing photons are produced before the blowout, which occurs slightly later at $z = 3$ than at $z = 8$. We conclude that the dynamic chimneys created by galactic outflows increase the escape of ionizing radiation in the high-redshift disks, in which otherwise only a negligible amount of photons can escape. With $F_{esc} \gtrsim 0.2$, dwarf galaxies may make a substantial contribution to reionizing the IGM in the high-redshift universe.

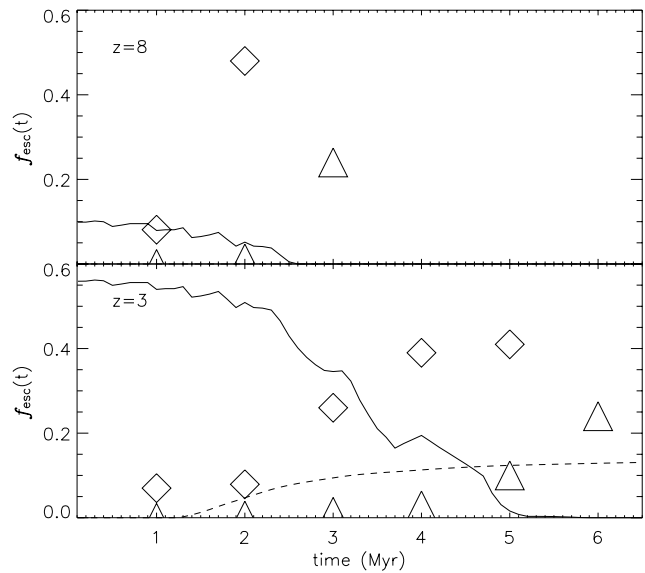

Fig. 13.—Instantaneous escape fractions $f_{\rm esc}(t)$ in undisturbed disks (line) and in disks with the effects of shells and galactic outflows (symbols) as a function of time. The models are disks with $M_d = 10^8\ M_\odot$ at $z = 3$ (bottom) and $z = 8$ (top), with $f_* = 0.06$. The Lyman continuum histories are based on the instantaneous-burst model, which yields $L_{\rm mech} = 10^{40}$ ergs s$^{-1}$ (solid line and diamonds), and on the multiple-burst model, which yields $L_{\rm mech} = 10^{39}$ ergs s$^{-1}$ (dashed line and triangles). The escape fractions $f_{\rm esc}(t)$ are computed up to when the bubbles blow out of the disks and galactic outflows are formed. Note that $f_{\rm esc}(t)$ increases once the outflows create funnels for the ionizing photons to escape directly to the IGM (except in the disk at $z = 3$ with the instantaneous-burst model).

### 5.2.4. Other Issues

We discussed only the results with $f_* = 0.06$. However, we can guess the trend of $f_{esc}(t)$ and $F_{esc}$ as a function of $f_*$, since the formation of dense swept-up shells due to isothermal shocks that trap ionizing radiation is expected for all the disks with mechanical luminosities corresponding to $f_* \leq 0.6$ at all redshifts, as long as the bubbles are within a few scale heights of the disks. The star formation efficiency $f_*$ has a linear dependence on $L_{mech}$, so the time for a blowout, $t_{bl} \propto L_{mech}^{-1/3}$, changes by only a factor of 2 with an order-of-magnitude increase in $f_*$ and $L_{mech}$. In addition, the photon luminosity $N_{ph}(t)$ has a linear dependence on $f_*$, while the effective recombination rate of the shells, $\Xi_{sh} \propto n_{sh} M^2$, is only a factor of ~5 larger with an order-of-magnitude increase in $f_*$, $N_{ph}(t)$, and $L_{mech}$ because $v_{sh} \propto L_{mech}^{1/3}$. Therefore, an early increase in $f_{esc}(t)$ and an increase in $F_{esc}$ are expected with $f_* = 0.6$. The opposite can be said of the behaviors of $f_{esc}(t)$ and $F_{esc}$ with $f_* = 0.006$. The summary of the effects of shells and galactic outflows on the total escape fractions $F_{esc}$ with different $f_*$ is listed in Table 4.

We did not run the simulations for large disks at high redshift, because the extremely strong shocks in high-density environments are unstable in ZEUS-3D. However, the extrapolation of our results indicates that the blowouts occur very quickly in high-redshift massive disks, so the total escape fraction is increased to $F_{esc} \gtrsim 0.2$. Note that the redshift dependence of $F_{esc}$ discussed in § 5.1 does not exist in the presence of galactic outflows.

We do not consider the effects of dust on the escape fractions of photons in our models. However, we think that the qualitative results will not change in our models even if realistic distributions of dust are included because the dust content is expected to be low in low-metallicity systems such as dwarf galaxies, especially at high redshift, and the dust will also be swept away with the galactic outflows. Note also that a hydrogen column density of $10^{21}$ cm$^{-2}$ is required for dust to produce $\tau_{dust} = 1$ even with solar metallicity (Spitzer 1978, eq. [7-23]), while an H I column of only $1.5 \times 10^{17}$ cm$^{-2}$ produces $\tau_{H I} = 1$ at the Lyman edge. Thus, dust can be neglected unless hydrogen is extremely ionized.

## 6. DISCUSSION

### 6.1. Comparison with Observations

In this study, we examined the effects of supershells and galactic outflows on the escape fraction from a single starburst cluster at the center of our model disk. Although this is only a first approximation to a realistic dwarf starburst galaxy, we can gain some insight by comparing our results with the observations of three dwarf starburst galaxies with escape fractions less than 0.06: NGC 1705, NGC 5253, and NGC 4214. In particular, observers measure the instantaneous escape fraction $f_{esc}(t)$, not the total escape fraction $F_{esc}$. The value of $f_{esc}(t)$ depends on starburst age and the stage of bubble evolution, as shown in the previous section, and therefore may not be representative of $F_{esc}$.

First, we go over the basic properties of the three dwarf starburst galaxies. Table 5 gives a summary.

1. NGC 1705 is a "nucleated" blue compact dwarf galaxy with multiple ellipsoidal shells expanding at $v_{sh} \sim 100$ km s$^{-1}$, driven by a super–star cluster NGC 1705-1 (Meurer, Freeman, & Dopita 1992; Meurer et al. 1995; Marlowe, Meurer, & Heckman 1999; Heckman et al. 2001). The ionizing continuum of the order of $10^{52}$ s$^{-1}$ is produced mainly by the background star-forming population, not by the nucleus, although it dominates the UV-optical flux from the galaxy. The nucleus has mass $\sim 1.5 \times 10^6\ M_\odot$ with an estimated age of ~13 Myr after a nearly instantaneous burst, and star formation rate in the background is 0.03 $M_\odot$ yr$^{-1}$ with a star formation timescale of ~1 Gyr (Meurer et al. 1992). The H I rotating disk shows a hint of disturbance with the optical/H I peaks offset from the H I dynamical center, but it is not clear whether the starburst was externally triggered (Meurer, Staveley-Smith, & Killeen 1998).

2. NGC 5253 is an amorphous dwarf galaxy with two Wolf-Rayet clusters. These clusters produce most of the observed ionizing radiation $\sim 4 \times 10^{52}$ s$^{-1}$ (Schaerer et al.

TABLE 5
Observed Properties of NGC 1705, NGC 5253, and NGC 4214

| Parameter | NGC 1705[a] | NGC 5253[b] | NGC 4214[b] |
|---|---|---|---|
| $d$ (Mpc) .......................... | 6.1 | 4.1 | 3.6 |
| $M_B$ (mag) ...................... | −16.20 | −17.62 | −17.65 |
| H I extent (kpc) .............. | ~4.8 | 4.3 × 4.3 | 14.6 × 12.9 |
| $v_{circ}$ (km s$^{-1}$) ................. | ~60 | 38–60 | 70 |
| $R(v_{circ})$ (kpc) ................ | 2.1 | 1.24 | 2.91 |
| Type............................... | Amorphous/BCD | Im(pec), amorphous/BCD | IAB(s)m |
| $M_{dyn}$ ($M_\odot$).................... | 1–2E9 | >4.3E8 | 3.3E9 |
| $M_{H I}$ ($M_\odot$).................... | 8.7E7 | 1.4E8 | 1.8E9 |
| $Q$ (s$^{-1}$)........................... | ~1E52 | 3.86E52 | 1.94E52 |
| $f_{esc}$ (%)........................... | <5.8 | <6.0 | <4.0 |

Note.—The escape fractions $f_{esc}(t)$ in these dwarf galaxies are estimated to be less than 0.06 (Heckman et al. 2001).

[a] The observed properties of NGC 1705 are taken from Meurer et al. 1992, 1998 and Marlowe et al. 1999.

[b] The observed properties of NGC 5253 and NGC 4214 are taken from Tables 1, 4, and 5 in Martin 1998.

1997) and drive superbubbles observed in H$\alpha$ probing filamentary structures expanding at $\sim$35 km s, coincident with soft X-ray emission (Martin & Kennicutt 1995; Martin 1998; Marlowe et al. 1999). The clusters are very young, with estimated ages 2.8 and 4.3 Myr after (nearly) instantaneous bursts and are embedded in an older star-forming background (0.1–1 Gyr). The measured stellar rotation rate of only less than 7 km s$^{-1}$ around the optical minor axis and the high stellar velocity dispersion (50 km s$^{-1}$) suggest that the galaxy was a dwarf elliptical before the current starburst began (Caldwell & Phillips 1989). An H I study of NGC 5253 shows that the bulk of the neutral hydrogen gas rotates about its optical major axis, and the peculiar H I kinematics may be a signature of recent accretion, disruption of tidal companions, or interaction with the nearby spiral galaxy M83 (Kobulnicky & Skillman 1995).

3. NGC 4214 is a (barred) irregular dwarf galaxy, also listed as a Wolf-Rayet galaxy. The total ionizing continuum of $3 \times 10^{52}$ s$^{-1}$ is mainly produced by several very young ($\sim$3 Myr) massive star clusters, which also drive superbubbles sweeping the ISM into shells with $v_{\rm sh} > 50$ km s$^{-1}$, creating two large H II complexes (Martin 1998; MacKenty et al. 2000). It has a very extended H I disk with a peak column density $\sim 3 \times 10^{21}$ cm$^{-2}$, and the overall H I distribution indicates the existence of H I holes and shells, also with a hint of a bar (McIntyre 1997; Walter et al. 2001). The total H I mass and estimated dynamical mass are larger than those of NGC 1705 and NGC 5253 (Martin 1998; see Table 5).

In NGC 5253 and NGC 4214, the ages of the superclusters are estimated to be $\lesssim 4$ Myr. As shown in Figure 11, the instantaneous escape fraction $f_{\rm esc}(t)$ is suppressed to less than $\sim$0.1 before the bubbles start to accelerate at $t \approx 4$ Myr in the thin disk or at $t \approx 10$ Myr in the thick disk with $M_d = 10^8\ M_\odot$ at $z = 0$. Our low-redshift disk models predict low escape fractions in NGC 5253 and NGC 4214 because of their young age. The observed kinematics support the hypothesis that the shells have not broken up yet. The low escape fraction measured in NGC 1705 requires a different explanation since the age of the supercluster is $\sim$13 Myr. However, multiple shells extending out to $\sim$1 kpc from the supercluster do not yet seem to be fragmenting because of Rayleigh-Taylor instability (Meurer et al. 1992). The low escape fraction in NGC 1705 is partly explained by our results that the blowout time is delayed until $\sim$10 Myr in thick disks with large scale heights.

Escape fractions have been observed for other galaxies: blue compact galaxies with $M_B > -20$ such as Mrk 66, Mrk 1267 (Leitherer et al. 1995; Hurwitz et al. 1997), and Mrk 54 (Deharveng et al. 2001), nucleated starburst galaxies, IRAS 08339+6517 and Mrk 496 (Leitherer et al. 1995; Hurwitz et al. 1997), Lyman break galaxies (Steidel et al. 2001), and high-redshift galaxies from the Hubble Deep Field (HDF) with magnitudes $M_{1500\ {\rm \AA}} \approx -21$ to $-25$ (Fernandez-Soto et al. 2003). The escape fractions from the low-redshift BCDs and starburst galaxies are estimated to be less than 10%. On the other hand, the escape fractions from LBGs at $z \sim 3.4$ from Steidel et al. (2001) are estimated to be greater than $\sim$50%, and those from HDF galaxies at $1.9 < z < 3.4$ from Fernandez-Soto et al. (2003) are estimated to be less than $\sim$15%. Although most of them are larger galaxies in which starburst activities extend to rather more than a few kiloparsecs (e.g., Giavalisco, Steidel, & Macchetto 1996), a simple extrapolation of our results with dwarf galaxies suggests that the escape fractions should correlate with the stages of bubble evolution and therefore with the ages of starbursts in the galaxies.

For the low-redshift BCDs and starburst galaxies, the ages of starbursts are estimated to be a few to $\sim$6 Myr, and these starburst clusters seem to drive large-scale motions of the interstellar gas at $v_{\rm sh}$ about several hundred km s$^{-1}$ (González Delgado et al. 1998). In light of our results, the swept-up shells around the young starburst bubbles may trap the ionizing photons, while the observed large-scale motions have not yet developed into galactic outflows. On the other hand, among 29 LBGs drawn from the bluest quartile of intrinsic UV colors, only two of them have their starburst ages estimated to be $\sim$1 Gyr under the assumption of constant star formation (Shapley et al. 2001). However, there seems to be a relationship such that LBGs with intrinsically bluer rest-frame UV colors, after correction for IGM absorption, are older starburst systems (Shapley et al. 2001). According to the results with our model disks with $M_d = 10^8\ M_\odot$ at high redshift, we could suggest that the escape fraction can easily be $\sim$0.5 in the LBGs, because the strong galactic outflows due to powerful starbursts can make chimneys allowing a significant number of ionizing photons to escape to the IGM. Many LBGs show evidence for large-scale galactic outflows (Pettini et al. 1998, 2001). However, Fernandez-Soto et al. (2003) do not find the trend of increasing escape fractions as a function of color of galaxies. The low escape fractions observed in them are hard to explain with our models, unless they are all starburst systems so young that the shells still trap the ionizing radiation.

### 6.2. UV Background Radiation

Our results suggest that a significant fraction of ionizing radiation escapes from high-redshift disks. The recent measurement by the *Wilkinson Microwave Anisotropy Probe* of a large electron scattering optical depth implies that a significant volume of the IGM was reionized by redshift $z \sim 17 \pm 5$ (Kogut et al. 2003; Spergel et al. 2003). Population III objects may account for such early reionization if efficient molecular hydrogen cooling occurs. However, the measured temperature of the Ly$\alpha$ forest at $z \sim 3$ rules out a single reionization model before $z \sim 9$ and suggests the reionization history is complex (Haiman & Holder 2003; Wyithe & Loeb 2003; Hui & Haiman 2003; Theuns et al. 2002). Therefore, we attempt to make an order-of-magnitude calculation of the contribution of escaping UV photons from dwarf starburst galaxies at $z > 5$ to the UV background radiation field and thus to the reionization of the universe.

We compute as a function of redshift the rest-frame comoving density of Lyman continuum photons escaping out of dwarf galactic halos. To do so, we first compute the formation rate of halos $R_{\rm form}$ and the survival probability of the halos $P_{\rm surv}$ as a function of redshift, using the excursion set halo mass function for a hierarchical random Gaussian density field (Bond et al. 1991) and the theory of hierarchical merging probabilities (Lacey & Cole 1994). Refer to Lacey & Cole (1994) for the detailed derivation of $R_{\rm form}$ and $P_{\rm surv}$. We use the spherical collapse model to define the critical density threshold for halo collapse and the CDM power spectrum given by Bardeen et al. (1986) with a scale-invariant initial condition to define the mass variance, as in Kitayama & Suto (1996). We employ a $\Lambda$CDM model with $\Omega_0 = 0.37$, $\Omega_\Lambda = 0.63$, $\Omega_b = 0.05$, $h = 0.7$, and $\sigma_8 = 0.8$, although $f_{\rm esc}(t)$ and $F_{\rm esc}$ were computed with our standard model (SCDM+$\phi_{\rm lcl}$) in the previous sections. Recall that the computation of $F_{\rm esc}$ is not sensitive to the change in cosmology nor dark-to-baryonic ratio within theoretical and observational uncertainties.

We select two groups of halos. First we look at halos with virial temperature $T_v > 10^4$ K [minimum halo mass $M_{\min}(z) = M_4(z)$], in which moderately efficient star formation is possible because of hydrogen line cooling. Second, we consider halos with $T_v > 10^5$ K or $v_c \geq 50$ km s$^{-1}$ [$M_{\min}(z) = M_5(z)$], in which $\gtrsim 50\%$ of the gas within the virial radius can cool in the reionized IGM (e.g., Thoul & Weinberg 1996). The maximum halo mass for the computation is set to $M_{\max} = 10^{11}\ M_\odot$. We assume the gas cools and settles into disks promptly at the redshift of halo collapse because the cooling time of the halos is much smaller than the Hubble time at high redshift, $t_{\rm cool} \ll t_H$ (Oh & Haiman 2002).

We then define the Lyman continuum photon background density due to dwarf starburst galaxies with star formation timescale $\tau_{\rm SF}$ as

$$\rho_{\rm UV} = \int_{M_{\min}(z)}^{M_{\max}} dM \int_z^\infty dz_f\, S(M, z_f, z) f_*$$
$$\times \left(\frac{\Omega_b}{\Omega_0} M\right) \frac{F_{\rm esc}(M, f_*, z) Q_{\rm ph}(M, f_*)}{\tau_{\rm SF}}, \quad (18)$$

where $M$ is the halo mass, $f_*$ is the star formation efficiency, assumed to be constant in $M$ and $z$, $Q_{\rm ph}(M, f_*)$ is the total number of photons produced per unit gram of gas converted to stars, and $S(M, z_f, z) dM\, dz_f = S(M, t_f, t) dM\, dt_f$ is the number density of halos that form with mass $M \sim M + dM$ at time $t_f \sim t_f + dt_f$ and survive without destruction until a later time $t$ (Lacey & Cole 1994). This can be expressed as

$$S(M, t_f, t) dM\, dt_f = R_{\rm form}(M, t) dM\, dt_f\, P_{\rm surv}(M, t_f, t). \quad (19)$$

Equations (18) and (19) simplify to

$$\rho_{\rm UV} \approx \int_{M_{\min}(z)}^{M_{\max}} f_* \left(\frac{\Omega_b}{\Omega_0} M\right)$$
$$\times F_{\rm esc}(M, f_*, z) Q_{\rm ph}(M, f_*) R_{\rm form}(M, z) dM \quad (20)$$

if we assume instantaneous starbursts, since most photons are produced within the first few Myr of the bursts. Equation (20) also reproduces equations (18) and (19) with multiple starbursts having $\tau_{\rm SF} = 20$ Myr because $R_{\rm form}$ and $P_{\rm surv}$ do not change over such a short timescale.

At $z > 5$, we assume $F_{\rm esc}(M, f_*, z)$ to be constant based on our results with $M_d = 10^8\ M_\odot$ at $z = 3$ and 8 with a star formation efficiency $f_* = 0.06$. To span the range of possibilities, we consider total escape fractions $F_{\rm esc} = 0.2$, 0.4, and 0.8, so that $f_* F_{\rm esc}$ takes on values of 0.012, 0.024, and 0.048. We also choose $f_* = 0.1$ and $F_{\rm esc} = 1.0$, so $f_* F_{\rm esc} = 0.1$ as an upper limit.

We show in Figure 14 the evolution of the UV background radiation as a function of redshift for $z > 5$ under our assumptions, compared with the minimum photon production rate needed to fully ionize the universe (i.e., when the ionization fronts completely overlap). The minimum rate is defined as $\mathcal{N} = \bar{n}_{\rm H\,I}(z=0)/\bar{t}_{\rm rec}$, where $\bar{n}_{\rm H\,I}(z=0)$ is the comoving mean hydrogen density of the expanding IGM and the volume-averaged recombination time $\bar{t}_{\rm rec} = [\bar{n}_{\rm H\,I}(z) \alpha_B C]^{-1}$ with the ionized hydrogen clumping factor $C$ (Madau et al. 1999; Gnedin & Ostriker 1997). This is based on the assumption that the clumping can be averaged over, and that the recombination time is very small, $\bar{t}_{\rm rec} \ll t_H(z)$ at high redshift so that only photons emitted within time $\bar{t}_{\rm rec}$ can actually ionize new material. We take values of $C$ ranging from 30 to 100 on the basis of numerical simulations (Gnedin 2000) and analytical studies (Benson et al. 2001; Haiman, Abel, & Madau 2001).

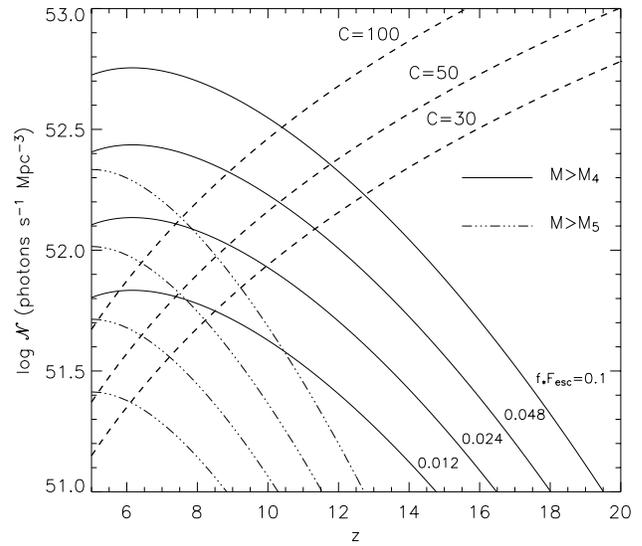

FIG. 14.—UV background density $\mathcal{N}$ (photons s$^{-1}$ comoving Mpc$^{-3}$) computed from two groups of halos (1) with $T_v > 10^4$ K ($M > M_4$; solid line) and (2) with $T_v > 10^5$ K ($M > M_5$; triple-dot–dashed line), based on the escape fractions $F_{\rm esc}$ in disks with $M_d = 10^8\ M_\odot$ at $z = 3$ and 8 studied above. The photon production rate is proportional to $f_* F_{\rm esc}$ when we assume that both $f_*$ and $F_{\rm esc}$ are constant in $z$ and $M_d$. It has to exceed the minimum photon production rate to fully ionize the universe if we assume clumping factors $C = 30$, 50, or 100 (dashed lines).

Figure 14 suggests that the Lyman continuum photon production rate from dwarf starburst galaxies are high enough to make a substantial contribution to reionize the IGM before $z \approx 10$ if $f_* F_{\rm esc} \gtrsim 0.024$; that is, $f_* \gtrsim 0.06$. We argued that $f_* \gtrsim 0.06$ is plausible on the basis of gravitational instability of our model disks and also with the presence of metals replenished from previous generations of stars. However, the formation of dwarf galaxies can be suppressed by mechanical feedback from other dwarf galaxies (Scannapieco, Thacker, & Davis 2001; Fujita et al. 2003). To study the process of reionization, we ultimately need three-dimensional cosmological simulations with radiative transfer, which can also resolve the formation of SN-driven bubbles and outflows. Our crude calculation still shows that dwarf starburst galaxies with $F_{\rm esc} \gtrsim 0.2$ with a reasonable star formation efficiency, $f_* \approx 0.05$–0.1, provide a substantial amount of UV photons to the IGM.

## 7. CONCLUSION

We used numerical models to study the influence of shells and galactic outflows on the escape of ionizing radiation from dwarf galactic disks with starburst clusters. From our models, we can draw the following conclusions:

1. The shells of ISM swept up by isothermal shocks around young superbubbles can trap ionizing radiation very

effectively until the bubbles start to accelerate, causing the shells to fragment. The acceleration of the bubbles begins when the bubbles expand to a few scale heights above the disks.

2. The low escape fractions of ionizing photons observed in local dwarf galaxies less than 0.06 can be explained if they are observed before blowout. However, the observed $f_{esc}(t)$ may not reflect the total fraction of photons escaping from the disks over the lifetime of the starbursts $F_{esc}$.

3. Galactic outflows blow out to form holes that allow ionizing radiation to escape directly to the IGM. The escape fractions are $F_{esc} \approx 0.2$ in the high-redshift disks, in which only a negligible amount of photons can escape without bubble dynamics. This is because the gas distributions are highly stratified in the strong gravitational potentials of dense, high-redshift halos, allowing quick acceleration of the bubbles.

4. With $f_* = 0.05$–$0.1$ and $F_{esc} \gtrsim 0.2$, dwarf starburst galaxies may have contributed significantly to the UV background radiation and so to the reionization of the IGM before $z \sim 10$.

The assumptions used in modeling dwarf galaxy disks and star formation distributions and histories can influence the predicted escape of ionizing radiation in several ways:

1. The vertical scale height $H$ influences the escape of ionizing photons because the recombination measure is $\Xi_Z \propto N_{H_I}^2$. Our thick undisturbed disks suppress escape more than our thin undisturbed disks. Galactic outflows enhance this effect because bubbles blow out of thin disks earlier than thick disks: $t_{bl} \propto H^{4/3}$. The time of blowout determines the transition from the trapping of photons by shells to their escape through chimneys in galactic disks.

2. In undisturbed disks, the peak photon luminosity primarily determines the escape fractions. The instantaneous escape fraction $f_{esc}(t)$ follows the behavior of $N_{ph}(t)$, predicted by a star formation history (e.g., instantaneous burst or multiple bursts). In disks with supershells and galactic outflows, $f_{esc}(t)$ strongly depends on the evolution of bubbles rather than on the star formation history. The escape fraction with a different star formation history from the ones we used in this study can be inferred on the basis of our results by computing and comparing its peak photon luminosity and expected mechanical luminosity.

3. The star formation efficiency $f_*$ we used in this study is for a single starburst cluster in a disk, not for the entire disk. When a fixed total star formation efficiency is given in the disk, the distribution of ionizing sources strongly influences the escape of ionizing photons because the photon luminosity from each ionizing source must exceed a minimum photon luminosity $N_{ph,min}$ for its Strömgren sphere to break out of the disk. A single ionizing source gives the most favorable condition for the photons to escape, compared with numerous OB associations scattered around the disk in space and in time, as considered in Dove & Shull (1994), Dove et al. (2000), Wood & Loeb (2000), and Ciardi et al. (2002).

Our models have several limitations:

1. Shell formation is not resolved even with resolutions as low as a few tenths of a parsec. Therefore, when we compute the escape of ionizing radiation in the simulations, we use the theoretical peak density to compute the recombination measure $\Xi$ along rays that encounter dense shells rather than holes produced by the Rayleigh-Taylor instability. This correction was necessary only in low-redshift disks with low hydrogen density.

2. The fragmentation of the shells is resolved qualitatively: the hot gas accelerates beyond the cold swept-up shells. The opening angles of chimneys cleared out by the Rayleigh-Taylor instability remain $30°$–$50°$ at the time of blowouts regardless of resolution. However, the quantitative details of the fragmentation, including the number and positions of individual fragments, are determined by our assumption of cylindrical symmetry, by numerical resolution, and by the lack of magnetic fields in our calculations.

3. We modeled only a single starburst cluster placed at the center of a rotationally supported disk. However, we argue that a starburst cluster placed at the center yields a lower limit for the escape fraction, which remains as a lower limit in a more realistic galaxy in which multiple star clusters of the same or smaller photon luminosities are scattered around the disk.

4. Our models include only a single-phase, smoothly stratified ISM. We argue that this yields a lower limit for the actual escape fraction for ionizing radiation.

5. We assume well-formed, rotationally supported disks, even at high redshift. A more turbulent ISM may form because of stellar feedback and cooling, perhaps even before the formation of a disk. Bubble blowout through such a medium is likely to be easier than through a well-formed disk. However, to study the escape fractions at high redshift and the reionization of the IGM, it will ultimately be necessary to follow the collapse of gas and solve the radiation transfer problem in three-dimensional cosmological simulations.

We thank A. Ferrara and A. Shapley for useful discussions and S. Kahn, Z. Haiman, and C. Schwartz for careful inspection of the manuscript. We also thank K. Nagamine for providing very useful data to us. A. F. was supported by the Beller Fellowship of the American Museum of Natural History and partly supported by the Cooperative Agreement for Research and Education between Los Alamos National Laboratory and the University of Santa Barbara. She thanks the California Institute of Technology and K. Wakamatsu at Gifu University for hospitality during work on this project. M.-M. Mac L. was partly supported by NSF CAREER grant AST 99-85392, and C. L. M. was partly supported by Smithsonian Astrophysical Observatory grant GO0-1140A. Computations were performed on the SGI Origin machines of the Hayden Planetarium, of the National Center for Supercomputing Applications, and of the Center for Advanced Computing Research at Caltech. We thank M. Norman and the Laboratory for Computational Astrophysics for use of ZEUS. This research made use of the Abstract Service of the NASA Astrophysics Data System.

REFERENCES

Abel, T., Anninos, P., Norman, M. L., & Zhang, Y. 1998, ApJ, 508, 518
Abel, T., Bryan, G. L., & Norman, M. L. 2000, ApJ, 540, 39
Abel, T., Norman, M. L., & Madau, P. 1999, ApJ, 523, 66
Bajtlik, S., Duncan, R. C., & Ostriker, J. P. 1988, ApJ, 327, 570
Bardeen, J. M., Bond, J. R., Kaiser, N., & Szalay, A. S. 1986, ApJ, 304, 15
Becker, R. H., et al. 2001, AJ, 122, 2850
Benson, A. J., Nusser, A., Sugiyama, N., & Lacey, C. G. 2001, MNRAS, 320, 153
Binney, J., & Tremaine, S. 1987, Galactic Dynamics (Princeton: Princeton Univ. Press)
Blondin, J. M., & Cioffi, D. F. 1989, ApJ, 345, 853
Bond, J. R., Cole, S., Efstathiou, G., & Kaiser, N. 1991, ApJ, 379, 440
Burkert, A. 1995, ApJ, 447, L25
Caldwell, N., & Phillips, M. M. 1989, ApJ, 338, 789
Castor, J., McCray, R., & Weaver, R. 1975, ApJ, 200, L107
Chevalier, R. A., & Imamura, J. N. 1982, ApJ, 261, 543
Ciardi, B., Bianchi, S., & Ferrara, A. 2002, MNRAS, 331, 463
Clarke, C., & Oey, M. S. 2002, MNRAS, 337, 1299
Clarke, D. A. 1994, NCSA Tech. Rep.
Cole, S., & Lacey, C. 1996, MNRAS, 281, 716
Dalcanton, J. J., Spergel, D. N., & Summers, F. J. 1997, ApJ, 482, 659
Deharveng, J.-M., Buat, V., Brun, V. L., Milliard, B., Kunth, D., Shull, J. M., & Gry, C. 2001, A&A, 375, 805
Della Ceca, R., Griffiths, R. E., Heckman, T. M., & MacKenty, J. W. 1996, ApJ, 469, 662
D'Ercole, A., & Brighenti, F. 1999, MNRAS, 309, 941
De Young, D. S., & Heckman, T. M. 1994, ApJ, 431, 598
Dickey, J. M., Hanson, M. M., & Helou, G. 1990, ApJ, 352, 522
Djorgovski, S. G., Castro, S., Stern, D., & Mahabal, A. A. 2001, ApJ, 560, L5
Donahue, M., & Shull, J. M. 1987, ApJ, 323, L13
Dove, J. B., & Shull, J. M. 1994, ApJ, 430, 222
Dove, J. B., Shull, J. M., & Ferrara, A. 2000, ApJ, 531, 846
Fan, X., et al. 2001, AJ, 121, 54
Fernandez-Soto, A., Lanzetta, K. M., & Chen, H.-W. 2003, MNRAS, 342, 1215
Franx, M., Illingworth, G. D., Kelson, D. D., van Dokkum, P. G., & Tran, K.-V. 1997, ApJ, 486, L75
Fujita, A., Mac-Low, M.-M., Ferrara, A., & Meiksin, A. 2003, ApJ, submitted
Gaetz, T. J., & Salpeter, E. E. 1983, ApJS, 52, 155
Giavalisco, M., Steidel, C. C., & Macchetto, F. D. 1996, ApJ, 470, 189
Gnedin, N. Y. 2000, ApJ, 535, 530
Gnedin, N. Y., & Ostriker, J. P. 1997, ApJ, 486, 581
González Delgado, R. M., Leitherer, C., Heckman, T., Lowenthal, J. D., Ferguson, H. C., & Robert, C. 1998, ApJ, 495, 698
Gunn, J. E., & Peterson, B. A. 1965, ApJ, 142, 1633
Haiman, Z., Abel, T., & Madau, P. 2001, ApJ, 551, 599
Haiman, Z., & Holder, G. P. 2003, ApJ, 595, 1
Haiman, Z., & Loeb, A. 1998, ApJ, 503, 505
Heckman, T. M., Lehnert, M. D., Strickland, D. K., & Armus, L. 2000, ApJ, 129, 493
Heckman, T. M., Sembach, K. R., Meurer, G. R., Leitherer, C., Calzetti, D., & Martin, C. L. 2001, ApJ, 558, 56
Hui, L., & Haiman, Z. 2003, ApJ, submitted (astro-ph/0302439)
Hunter, D. A., Elmegreen, B. G., & Baker, A. L. 1998, ApJ, 493, 595
Hunter, D. A., Gallagher, J., & Rautenkranz, D. 1982, ApJS, 49, 53
Hurwitz, M., Jelinsky, P., & Dixon, W. V. D. 1997, ApJ, 481, L31
Kennicutt, R. C., Jr. 1989, ApJ, 344, 685
———. 1998, ApJ, 498, 541
Kitayama, T., & Suto, Y. 1996, ApJ, 469, 480
Kobulnicky, H. A., & Skillman, E. D. 1995, ApJ, 454, L121
Kogut, A., et al. 2003, ApJS, 148, 161
Kompaneets, A. S. 1960, Soviet Phys. Dokl., 5, 46
Lacey, C., & Cole, S. 1994, MNRAS, 271, 676
Lehnert, M. D., & Heckman, T. M. 1996, ApJ, 462, 651
Leitherer, C., Ferguson, H. C., Heckman, T. M., & Lowenthal, J. D. 1995, ApJ, 454, L19
Leitherer, C., et al. M. 1999, ApJS, 123, 3
MacDonald, J., & Bailey, M. E. 1981, MNRAS, 197, 995
MacKenty, J. W., Maíz-Apellániz J., Pickens, C. E., Norman, C. A., & Walborn N. R. 2000, AJ, 120, 3007
Mac Low, M.-M., & Ferrara, A. 1999, ApJ, 513, 142 (MF99)
Mac Low, M.-M., & McCray, R. 1988, ApJ, 324, 776
Mac Low, M.-M., McCray, R., & Norman, M. L. 1989, ApJ, 337, 141
Mac Low, M.-M., & Norman, M. L. 1993, ApJ, 407, 207
Madau, P., Haardt, F., & Rees, M. J. 1999, ApJ, 514, 648
Marlowe, A. T., Heckman, T. M., Wyse, R. F. G., & Schommer, R. 1995, ApJ, 438, 563
Marlowe, A. T., Meurer, G. R., & Heckman, T. M. 1999, ApJ, 522, 183
Martin, C. L. 1998, ApJ, 506, 222
———. 1999, ApJ, 513, 156
Martin, C. L., & Kennicutt, R. C., Jr. 1995, ApJ, 447, 171
———. 2001, ApJ, 555, 301
Mateo, M. 1997, in ASP Conf. Ser. 116, The Second Stromlo Symposium: The Nature of Elliptical Galaxies, ed. M. Arnaboldi, G. S. Da Costa, & P. Saha (San Francisco: ASP), 259
McIntyre, V. J. 1997, Publ. Astron. Soc. Australia, 14, 122
McKee, C. F., van Buren, D., & Lazareff, B. 1984, ApJ, 278, L115
Meurer, G. R., Freeman, K. C., Dopita, M. A., & Cacciari, C. 1992, AJ, 103, 60
Meurer, G. R., Heckman, T. M., Leitherer, C., Kinney, A., Robert, C., & Garnett, D. R. 1995, AJ, 110, 2665
Meurer, G. R., Staveley-Smith, L., & Killeen, N. E. B. 1998, MNRAS, 300, 705
Mo, H. J., Mao, S., & White, S. D. M. 1998, MNRAS, 295, 319
Moos, H. W., et al. 2000, ApJ, 538, L1
Navarro, J. F., Frenk, C. S., & White, S. D. M. 1997, ApJ, 490, 493
Oh, S. P., & Haiman, Z. 2002, ApJ, 569, 558
Persic, M., Salucci, P., & Stel, F. 1996, MNRAS, 281, 27
Pettini, M., Shapley, A. E., Steidel, C. C., Cuby, J.-G., Dickinson, M., Moorwood, A. F. M., Adelberger, K. L., & Giavalisco, M. 2001, ApJ, 554, 981
Pettini, M., Kellogg, M., Steidel, C. C., Dickinson, M., Adelberger, K. L., & Giavalisco, M. 1998, ApJ, 508, 539
Ricotti, M., & Shull, J. M. 2000, ApJ, 542, 548
Salpeter, E. E. 1955, ApJ, 121, 161
Salucci, P., & Persic, M. 1997, in Dark and Visible Matter in Galaxies, ed. M. Persic & P. Salucci (San Francisco: ASP), 1
Scannapieco, E., Thacker, R. J., & Davis, M. 2001, ApJ, 557, 605
Schaerer, D., Contini, T., Kunth, D., & Meynet, G. 1997, ApJ, 481, L75
Schmidt, M. 1959, ApJ, 129, 243
Shapiro, P. A., & Giroux, M. L. 1987, ApJ, 321, L107
Shapley, A. E., Steidel, C. C., Adelberger, K. L., Dickinson, M., Giavalisco, M., & Pettini, M. 2001, ApJ, 562, 95
Shull, M. J., & Saken, J. M. 1995, ApJ, 444, 663
Silich, S. A., & Tenorio-Tagle, G. 1998, MNRAS, 299, 249
———. 2001, ApJ, 552, 91
Skillman, E. D. 1997, Observatory, 117, 313
Skillman, E. D., Kennicutt, R. C., & Hodge, P. W. 1989, ApJ, 347, 875
Spergel, D., et al. 2003, ApJS, 148, 175
Spitzer, L. 1978, Physical Processes in the Interstellar Medium (New York: Wiley)
Steidel, C. C., Adelberger, K. L., Giavalisco, M., Dickinson, M., & Pettini, M. 1999, ApJ, 519, 1
Steidel, C. C., Giavalisco, M., Pettini, M., Dickinson, M., & Adelberger, K. L. 1996, ApJ, 462, L17
Steidel, C. C., Pettini, M., & Adelberger, K. L. 2001, ApJ, 546, 665
Stone, J. M., & Norman, M. L. 1992, ApJS, 80, 753
Suchkov, A. A., Balsara, D. S., Heckman, T. M., & Leitherer, C. 1994, ApJ, 430, 511
Suchkov, A. A., Berman, V. G., Heckman, T. M., & Balsara, D. S. 1996, ApJ, 463, 528
Sutherland, R. S., & Dopita, M. A. 1993, ApJS, 88, 253
Swaters, R. A., van Albada, T. S., van der Hulst, J. M., & Sancisi, R. 2002, A&A, 390, 829
Theuns, T., Schaye, J., Zaroubi, S., Tae-Sun, K., Tzanavaris, P., & Carswell, B. 2002, ApJ, 567, L103
Thoul, A. A., & Weinberg, D. H. 1996, ApJ, 465, 608
Tomisaka, K., & Ikeuchi, S. 1988, ApJ, 326, 208
Toomre, A. 1963, ApJ, 138, 385
Wada, K., & Venkatesan, A. 2003, ApJ, 591, 38
Walter, F., Taylor, C. L., Hüttemeister, S., Scoville, N., & McIntyre, V. 2001, AJ, 121, 727
Weaver, R., McCray, R., Castor, J., Shapiro, P., & Moore, R. 1977, ApJ, 218, 377
Wood, K., & Loeb, A. 2000, ApJ, 545, 86
Wyithe, J. S. B., & Loeb, A. 2003, ApJ, 588, 69
Vacca, W. D. 1996, in The Interplay Between Massive Star Formation, the ISM and Galaxy Evolution, ed. D. Kunth, B. Guiderdoni, M. Heydari-Malayeri, & T. X. Thuan (Gif-sur-Yvette: Editions Frontières), 321
van der Kruit, P. C., & Shostak, G. S. 1984, A&A, 134, 258
van Leer, B. 1977, J. Comput. Phys., 23, 276
van Zee, L., Skillman, E. D., & Salzer, J. J. 1998, AJ, 116, 1186
Vishniac, E. T. 1983, ApJ, 274, 152
Yabe, T., & Xiao, F. 1993, J. Phys. Soc. Japan, 62, 2537